\documentclass[preprint]{aastex701}
\usepackage{CJK}
\usepackage{amsmath}
\graphicspath{{./}{figures/}}

\begin{document}
\begin{CJK*}{UTF8}{gbsn}
\title{New $H(z)$ measurement at Redshift = 0.12 with DESI Data Release 1}

\author[orcid=0009-0004-9366-1947,sname='Wang']{Ze-fan Wang(王泽凡)}
\affiliation{Key Laboratory of Dark Matter and Space Astronomy, Purple Mountain Observatory, Chinese Academy of Sciences, Nanjing 210023, China}
\affiliation{School of Astronomy and Space Science, University of Science and Technology of China, Hefei 230026, China}
\email[show]{zfwang@pmo.ac.cn} 
\correspondingauthor{Ze-fan Wang(王泽凡), Lei Lei(雷磊)}

\author[orcid=0000-0003-4631-1915]{Lei Lei(雷磊)} 
\affiliation{Key Laboratory of Dark Matter and Space Astronomy, Purple Mountain Observatory, Chinese Academy of Sciences, Nanjing 210023, China}
\affiliation{School of Astronomy and Space Science, University of Science and Technology of China, Hefei 230026, China}
\email[show]{leilei@pmo.ac.cn}

\author[0000-0002-8966-6911,sname='Fan']{Yi-zhong Fan(范一中)}
\email{yzfan@pmo.ac.cn}
\affiliation{Key Laboratory of Dark Matter and Space Astronomy, Purple Mountain Observatory, Chinese Academy of Sciences, Nanjing 210023, China}
\affiliation{School of Astronomy and Space Science, University of Science and Technology of China, Hefei 230026, China}

\begin{abstract}
The Hubble parameter ($H(z)$) is a function of the redshift and a reliable measurement is very important to understand the expansion history of the Universe. In this work, we perform full-spectrum fitting using \texttt{BAGPIPES} on more than four thousand massive, passively evolving galaxies released by the DESI collaboration to estimate their cosmological-independent stellar ages and star-formation histories, and derive a new measurement of $H(z=0.12)=71.33 \pm 4.20~{\rm km~s^{-1}~Mpc^{-1}}$, which is well consistent with those derived in other ways.

\end{abstract}



\section{Introduction}

Modern cosmology is built upon the $\Lambda$CDM framework, which features a cosmological constant ($\Lambda$) and cold dark matter (CDM). The cosmological constant $\Lambda$ drives the accelerated expansion of the Universe, while CDM plays a central role in the formation of galaxies and large-scale structure. This paradigm has been firmly established through multiple cosmological probes, including the cosmic microwave background (CMB; \citealt{1992ApJ...396L...1S, 2020A&A...641A...6P}), Type~Ia supernovae \citep{1998AJ....116.1009R, 2022ApJ...938..110B}, and baryon acoustic oscillations (BAO; \citealt{2001MNRAS.327.1297P, 2005ApJ...633..560E, 2025PhRvD.112h3515A}). Within this model, the Hubble constant is a fundamental observable, directly related to the present expansion rate of the Universe and its inferred age. However, the increasing precision of these measurements has revealed a significant tension in the inferred value of this parameter \citep{2022JHEAp..34...49A, 2023ARNPS..73..153K}. It is therefore crucial to develop and employ independent probes \citep{2022LRR....25....6M} to determine whether this discrepancy arises from unresolved systematic effects or signals new physics beyond the standard cosmological model \citep{2021CQGra..38o3001D}.

Under the assumption of a Friedmann--Lema\^{\i}tre--Robertson--Walker (FLRW) metric, the Hubble parameter can be expressed as a function of redshift,
\begin{equation} \label{H(z)}
H(z) = -\frac{1}{1 + z} \frac{dz}{dt},
\end{equation}
where $z$ denotes redshift and $t$ is the cosmic time. A straightforward approach to addressing the Hubble tension is to extrapolate $H(z)$ to zero redshift. Accordingly, two main methods have been proposed to measure the expansion history of the Universe: one based on the radial BAO scale \citep{2003astro.ph..1623E, 2003ApJ...594..665B}, and the other relying on the differential galaxies' age method \citep{2002ApJ...573...37J}. The former exploits the inverse relationship between $H(z)$ and the differential radial comoving distance, which can be inferred from the radial extent of BAO features across redshift. However, this technique requires a prior calibration of the comoving BAO scale, typically derived from CMB observations. As a result, BAO-based measurements are not strictly cosmological model-independent, since the sound horizon scale is usually inferred within a specific cosmological framework \citep{2020A&A...641A...6P}.

The second approach determines the Hubble parameter by measuring the differential age--redshift relation of galaxies. In this case, $H(z)$ is directly calculated from the local derivative of cosmic age ($dt$) with respect to redshift ($dz$), and systematic offsets associated with absolute age measurements are largely mitigated. Therefore, the careful selection of a class of galaxies that can serve as reliable ``clocks'' is crucial for this method. \citet{2002ApJ...573...37J} proposed that massive and passively evolving galaxies are optimal candidates for such ``cosmic chronometers'' (CCs). Numerous studies have shown that these galaxies formed the bulk of their stellar mass rapidly and at high redshift \citep{1996AJ....112..839C,2011JCAP...03..045M, 2014ApJ...788...72G, 2019MNRAS.490..417C, 2022ApJ...929..131C}, resulting in a homogeneous and synchronized stellar population that effectively acts as a cosmic chronometer \citep{1996AJ....112..839C, 2010MNRAS.404.1775T, 2018MNRAS.480.4379C}. Alternatively, the age of quasars with higher redshifts can also be used to calculate the lower limit of $H(z)$ \citep{2022JHEAp..36...27V}. To construct a pure CC sample, a variety of selection criteria have been employed, including galaxy morphology \citep{hubble1982realm}, color--color cuts \citep{2013A&A...556A..55I, 2012JCAP...08..006M, 2016JCAP...05..014M}, spectral energy distribution (SED) fitting \citep{2010A&A...523A..13P}, and spectroscopic features such as Lick indices \citep{2018ApJ...868...84M, 2022ApJ...927..164B}. Comparisons among different selection strategies indicate that a relatively mature and robust procedure for constructing CC samples has been established \citep{2013A&A...558A..61M, 2018ApJ...868...84M, 2022LRR....25....6M}. 

Determining accurate galaxy ages, however, remains challenging. A number of approaches have been explored, including full-spectrum fitting \citep{2005PhRvD..71l3001S, Zhang_2014, 2023A&A...679A..96T, 2023ApJS..265...48J} and methods based on specific spectral regions or indices \citep{2012JCAP...08..006M, 2016JCAP...05..014M, 2022ApJ...928L...4B}. As a largely model-independent probe, CC-based measurements of $H(z)$ provide a robust dataset for testing cosmological models and constraining the properties of dark energy \citep{2023JCAP...11..051E, 2026MNRAS.547ag430L, 2025arXiv250819081W}.

In this work, we make use of Data Release~1 (DR1) from the Dark Energy Spectroscopic Instrument (DESI), together with the DESI Legacy Imaging Surveys, which provide an extensive repository of high-quality spectroscopic and photometric data. We perform full-spectrum fitting using \texttt{BAGPIPES}\footnote{\href{https://bagpipes.readthedocs.io}{Bayesian Analysis of Galaxies for Physical Inference and Parameter EStimation (BAGPIPES): https://bagpipes.readthedocs.io}} \citep{2018MNRAS.480.4379C} on more than four thousand massive, passively evolving galaxies to estimate their stellar ages and star-formation histories, and derive a new measurement of $H(z)$ at low redshift. This paper is organized as follows. In Section~\ref{sec: data}, we describe the data and sample selection. In Section~\ref{sec: method}, we present the methodology and details of the full-spectrum fitting procedure. In Section~\ref{result} we show our properties of selected sample and discuss the $H(z)$ measurements. The last section concludes the whole paper. 

\section{Data} \label{sec: data}

\subsection{DESI Survey}

DESI is a powerful spectrometric facility mounted on the Nicholas U. Mayall 4-meter Telescope at Kitt Peak National Observatory in Arizona \citep{2022AJ....164..207D,2023AJ....166..259S,2023AJ....165..144G}. DESI employs 5,000 robotic fiber positioners over a 3.2-degree field of view, enabling the simultaneous acquisition of thousands of spectra. The instrument provides continuous wavelength coverage from 3600 to 9800 \text{\AA} with a spectral resolution of $R \sim 2000-5000$ \citep{2024AJ....168..245P,2023AJ....165....9S,2024AJ....168...95M}. DESI DR1 marks a monumental milestone for the project, serving as the first major public release of data from the Main Survey operations \citep{2025arXiv250314745D}. Covering observations taken between May 2021 and June 2022, DR1 presents a large catalog of extragalactic spectroscopy. The release contains high-quality spectra for over 18 million unique targets, including approximately 13 million galaxies and 1.6 million quasars, alongside roughly 4 million stars from our own Milky Way.

DESI spectroscopic targets are selected primarily from the DESI Legacy Imaging Surveys, which provide uniform optical ($g$, $r$, $z$) photometry over $\sim$20,000 deg$^{2}$, complemented by mid-infrared data ($W1,\, W2,\, W3$ and $W4$) from Wide-field Infrared Survey Explorer \citep{2017PASP..129f4101Z, 2019AJ....157..168D}. The Legacy Surveys are approximately two magnitudes deeper than Sloan Digital Sky Survey and form the photometric foundation for DESI target selection, ensuring a well-defined and homogeneous parent sample for statistical analyses. 

As our work intends to do, the galaxy sample can be used to measure the precise value of $H(z)$. All data and Value-Added Catalogs (VACs)\footnote{\url{https://data.desi.lbl.gov/doc/releases/dr1}} used in this work are now publicly available. We retrieve the data from SPectra Analysis and Retrievable Catalog Lab (SPARCL) and the Astro Data Lab \citep{2014SPIE.9149E..1TF, 2020A&C....3300411N}. There is \textbf{an} issue\footnote{\url{https://data.desi.lbl.gov/doc/releases/dr1/known-issues/}} in flux uncertainty estimation of DESI DR1 data and we use a step function to fix this. 

\subsection{Sample Selection}

In this study, we utilize the Stellar Mass and Emission Line Catalog (\cite{2024ApJ...961..173Z} and Zou Hu et al. (2025) in preparation). This catalog provides the stellar mass and emission line measurements for all galaxies in DESI DR1 with reliable redshift measurements. Stellar mass is \added{derived} by using \texttt{CIGALE} \citep{2009A&A...507.1793N, 2011ApJ...740...22S}, which employs the broad-band $g, r, z, W1$, and $W2$ photometry from the DESI Legacy Imaging Surveys, and spectrophotometry of 10 artificial bands generated through convolution with DESI spectra. Main optical emission lines are measured by a single Gaussian fit, with absorption correction through continuum fitting performed by \texttt{STARLIGHT} \citep{2005MNRAS.358..363C, 2007MNRAS.375L..16C}. Additionally, the catalog includes stellar population properties derived by \texttt{CIGALE} and those obtained by \texttt{STARLIGHT} using DESI spectra.

Particularly, we perform the sample selection by following steps:
\begin{enumerate}
    \item \textbf{Galaxy morphology selection}: The catalog provides the morphological type of galaxies and we specially select the early-type galaxies, namely \texttt{MORPHTYPE == DEV} in the catalog. 
    \item \textbf{Mass and specific star formation rate (sSFR) cut}: Next, we select stellar mass $\log_{10}(M_*/M_\odot) > 11$ and constrain the sSFR ($\rm{sSFR = Star~Formation~Rate / M_*}$), $\log_{10}(\rm sSFR) < -11$ to select quiescent galaxies. 
    \item \textbf{Emission line cut}: We clean the sample with obvious emission \textbf{lines}. We choose the equivalent width (EW) of [O\,{\footnotesize II}], [H\,{\footnotesize $\beta$}], [H\,{\footnotesize $\alpha$}] $<$ 4 \text{\AA} and the signal-to-noise ratio (S/N) of these emission line $< 3$. 
    \item \textbf{H/K cut}: Another diagnostic is the ratio of two absorption lines, CaII H at 3969 \text{\AA} and CaII K at 3934 \text{\AA}. We measure two pseudo-Lick indices with \texttt{pyLick}\footnote{\href{https://pylick.readthedocs.io/}{pyLick: https://pylick.readthedocs.io/}} \citep{2022ApJ...927..164B} and select the galaxies with $H/K < 1.3$. 
    \item \textbf{Redshift cut}: In order to fully use the DESI spectra and select high S/N sample, we decide to cut the redshift of our sample $z < 0.2$ and the key word \texttt{FLUX SCALE} in the catalog $> 3$. 
\end{enumerate}

\section{Method} \label{sec: method}

We utilize a modified version of the public code \texttt{BAGPIPES} whose cosmological prior is removed (detailed modification in sec.3 of \cite{2023ApJS..265...48J}) to fit jointly spectra and photometry with a Bayesian approach\citep{2019MNRAS.490..417C}. 

\subsection{Basics of \texttt{BAGPIPES}}

\texttt{BAGPIPES} models observed spectra and photometry by forward--modeling synthetic observables over a parameter space and inferring posterior distributions through Bayesian inference. The exploration of the parameter space and the estimation of posterior probabilities are performed using a nested sampling algorithm, either \texttt{MultiNest}\footnote{\url{https://johannesbuchner.github.io/PyMultiNest/}}\citep{2019OJAp....2E..10F, 2014A&A...564A.125B} or \texttt{Nautilus}\footnote{\url{https://nautilus-sampler.readthedocs.io/en/stable/}}\citep{2023MNRAS.525.3181L}, which efficiently samples high--dimensional and potentially degenerate parameter spaces. \added{In this research, we use the \texttt{Nautilus} sampling algorithm and its default setting by \texttt{BAGPIPES}}

The physical model implemented in \texttt{BAGPIPES} consists of these main components:

\begin{itemize}
    \item Stellar population synthesis (SPS).
    The intrinsic stellar emission is modeled using the 2016 version of the \cite{2003MNRAS.344.1000B} stellar population synthesis models, constructed assuming a \cite{2001MNRAS.322..231K} initial mass function. 

    \item Star formation history (SFH). 
    The star formation history, $\mathrm{SFR}(t)$, describes the temporal build--up of stellar mass and sets the relative contribution of SSPs of different ages. \texttt{BAGPIPES} allows for flexible parametric forms of SFH, enabling both rapid early formation and extended star formation scenarios. The specific forms adopted in this work are described in Section~\ref{subsubsec: SFH}.

    \item Interstellar medium (ISM) transmission. For the ionized phase, we follow the prescriptions of \cite{2001MNRAS.323..887C} to account for nebular emission and continuum associated with recent star formation. This optional component is included to conservatively capture potential residual ionized gas contributions. Regarding the neutral ISM, dust attenuation is implemented via wavelength-dependent transmission functions that modify the intrinsic stellar spectrum, allowing for the adoption of various attenuation laws.
    

\end{itemize}

In addition to these physical components, \texttt{BAGPIPES} provides optional non--physical modules to account for residual calibration mismatches between spectra and photometry, as well as additional noise terms to correct for underestimated observational uncertainties. These settings and their implementation are described in detail in Section~\ref{fitting}.

\subsubsection{SFH choice} \label{subsubsec: SFH}

The traditional approach to SED fitting has been to use a simple functional form to parameterize the SFH. In this work, we choose 2 commonly used forms to reconstruct the SFHs of CC:
\begin{enumerate}
\item The delayed exponentially declining (DED) SFH, which can be expressed as
\begin{equation} \label{delayed}
\mathrm{SFR}(t) \propto \begin{cases}
(t - T_0)e^{-\frac{t - T_0}{\tau}} & t \geq T_0\\
0 & t < T_0
\end{cases},
\end{equation}
where $T_0$ represents the time when star formation begins and $\tau$ is the timescale after which SFR declines exponentially. It is a typical form for quiescent galaxy because it is characterized by a burst episode of star formation followed by passive evolution \citep{2017MNRAS.469.3108C}.
\item The double-power-law (DPL) SFH, which can be expressed as 
\begin{equation} \label{dblplaw}
\mathrm{SFR}(t) \propto \left[ \left(\frac{t}{\tau}\right)^{\alpha} + \left(\frac{t}{\tau}\right)^{-\beta} \right]^{-1},
\end{equation}
where $\alpha$ and $\beta$ are the falling and rising slopes respectively, and $\tau$ is also related to the time at which star formation peaks. It is worthwhile to note that \cite{2023A&A...679A..96T} found the decoupling between the rising and falling slopes may lead to non-physical SFHs. To address this, we impose a conservative prior on $\beta$ in the range 10 to 1000, and assign $\alpha$ a logarithmic prior between 0.1 and 1000. The parameter $\tau$ is allowed to vary between 0 and 20 in a uniform prior.
\end{enumerate}
It is clear that there is a beginning of star formation in Equation \ref{delayed}. The definition of galaxies' age can be directly written as 
\begin{equation}
age = t_U(z_{obs}) - T_0, 
\end{equation}
where $t_U(z_{obs})$ is the age of the universe at the redshift of observation. On the other hand, Equation \ref{dblplaw} has no obvious  beginning nor the end of star formation. Hence, \texttt{BAGPIPES} offer \textbf{an} alternative age named \textit{mass--weighted age}, which is expressed as 
\begin{equation}
age_{mw}=t_U(z_{obs})-\frac{\int_0^{t_{obs}}t\mathrm{SFR(t)}dt}{\int_0^{t_{obs}}\mathrm{SFR(t)}dt}.
\end{equation}
The \textit{mass-weighted age} provides a robust estimate of the typical stellar age in a galaxy, as $T_0$ is modified by averaging the SFR(t) of galaxy, giving higher importance to those that contribute most to the total stellar mass. Thus, it primarily represents the time when the main bulk of star formation occurred. When using the Equation \ref{dblplaw} as SFH parameterization, the galaxy age provided by \texttt{BAGPIPES} is the \textit{mass-weighted age}. 

\subsection{fitting with DESI spectra} \label{fitting}

\begin{figure*}[ht!]
\plotone{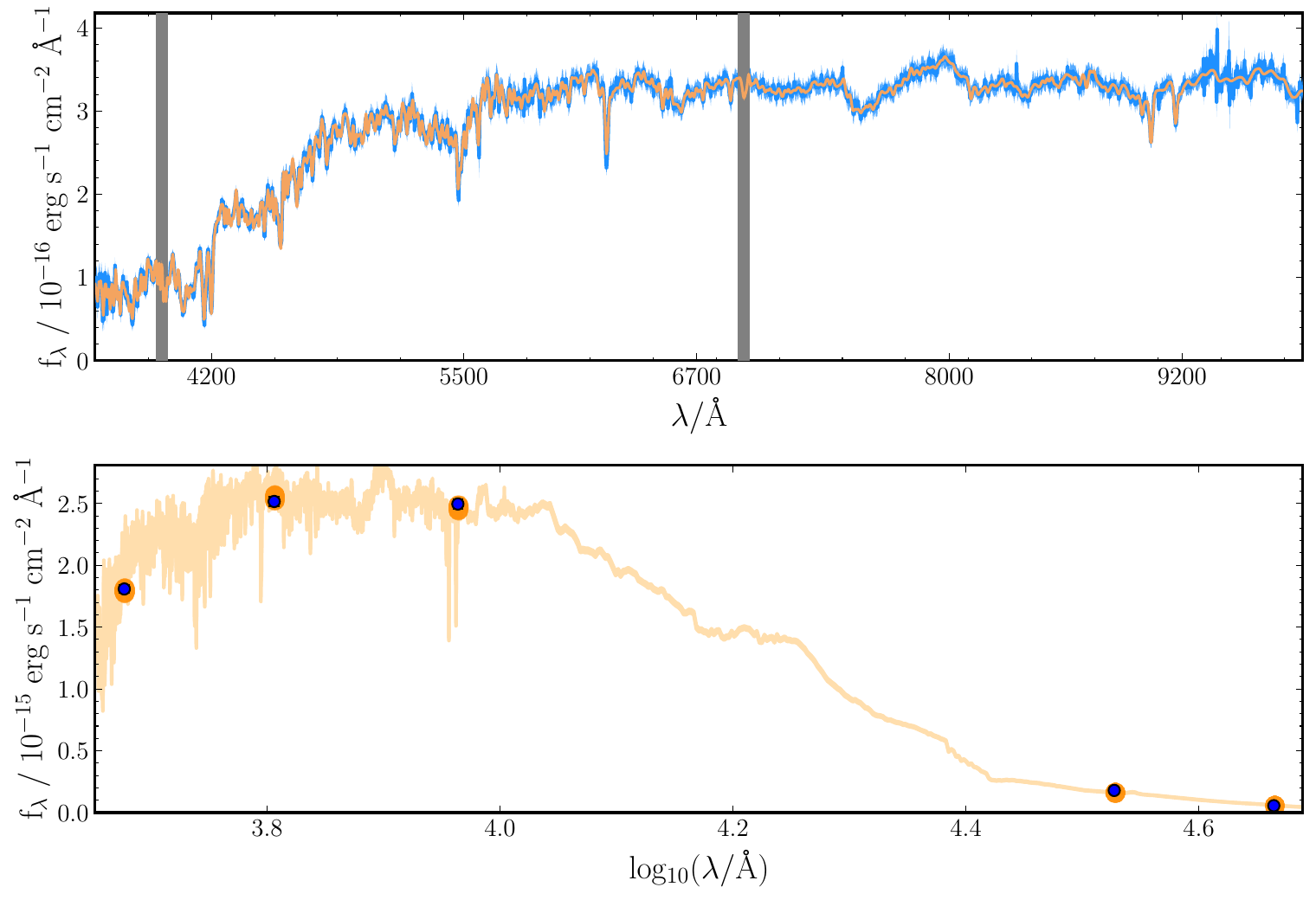}
\caption{\added{A typical CC (\texttt{TARGETID} = 2851244993413120) fitting result. Spectrum and photometry observation data are shown in blue, best-fit results are shown in orange. Gray areas are masked when fitting. }
\label{fig:CC}}
\end{figure*}

Before fitting the data, we check the potential anomalies in observation, finding an issue similar to \cite{2023A&A...679A..96T}. The S/N of the photometry is far larger than that of spectrum. To fix this circumstance, we assume the S/N for the photometry to be twice larger than the S/N of each spectrum\footnote{In this context, the S/N of a spectrum is defined as the median of the whole spectrum.}. In this way, we balance these two observations and also, the DESI DR1 spectra were binned by a factor of four to speed up the fitting as well as reduce the noise. 

We tested two fitting configurations whose major difference is the choices of SFH. The left identical components are listed below: 
\begin{itemize}
    \item a dust component described by \cite{2000ApJ...533..682C}. Regarding the stellar continuum attenuation, we impose a uniform prior on the $V$-band optical depth within the range $0 < A_V < 8$. The attenuation factor is fixed at $\epsilon = 2$, representing the ratio of dust attenuation in stellar birth clouds relative to the diffuse ISM. Additionally, we set the lifetime of these birth clouds to $t_{BC} = 10$ Myr; consequently, emission from stars formed within the last 10 Myr is subject to twice the attenuation experienced by the older population; 

    \item a nebular emission component described by \cite{2018MNRAS.480.4379C} using the \texttt{CLOUDY} \citep{2017RMxAA..53..385F}. Our sample selection should have already excluded this type of objects, but we include it in the model due to a more realistic physical simulation as well as an additional check of its absence. We fix the ionization parameter to $\log_{10}(U) = -3$ and mask the rest-frame region from 3702--3752 \text{\AA}, 6535--6585 \text{\AA} due to the ambiguities of low-level emission in quiescent galaxies. These two areas contain the [O\,{\footnotesize II}] and [H\,{\footnotesize $\alpha$}] emission line. AGN and ionization from old stars both thought to contribute to this phenomenon \citep{2013A&A...558A..43S, 2018MNRAS.481.1774H}. 

    \item a Gaussian process noise model described in \cite{2019MNRAS.490..417C}. Normally, the flux uncertainty of spectra is treated as uncorrelated or white noise. However, this hypothesis does not work when facing consecutive spectra pixels, due to many data-reduction steps from the raw spectroscopic observations into 1D spectral arrays \citep{2003MNRAS.343.1145P, 2019ApJ...874...17B}. Thus, a noise model based on Gaussian process is proposed, which used a covariance matrix describing the covariance between two wavelength bins:
    \begin{equation} \label{GP}
    \mathrm{C}_{jk}(\Phi)=a^{2}\sigma_{j}\sigma_{k}\delta_{jk}+b^{2}\mathrm{exp}{\left(-\frac{(\lambda_{j}-\lambda_{k})^{2}}{2l^{2}}\right)}
    \end{equation}
    where $\mathrm{C}_{jk}(\Phi)$is the covariance matrix, $\Phi$ is the nuisance parameters, $\sigma_{j,k}$ are the uncertainties on our pixel fluxes, $\lambda_{j,k}$ are the central wavelengths of our pixels, $\delta_{j,k}$ is the Kronecker delta function, and $a, b$ and $l$ are free parameters. The first term of Equation \ref{GP} deals with the uncorrelated noise and the second term represents the correlation between pixels. \added{Appendix \ref{apx: B} shows a more detailed for this model.} 
    
    \item a second-order Chebyshev polynomial calibration component \citep{2019MNRAS.490..417C}. This is designed for potential flux loss owing to a bunch of issues like the aperture correction and mismatching between the models and data. We apply Gaussian priors to the first and second polynomial coefficients with standard deviations of $\sigma =$ 0.25. The prior means are $\mu = $ 0 for the first and second order. For the zero order, we apply an uniform prior from 2 to 50 according to the \texttt{FLUX SCALE} in the DESI catalog for aperture correction. 
\end{itemize}
The parameter space and prior form are concluded in Table \ref{tab:para}. 

\begin{figure*}[ht!]
\plotone{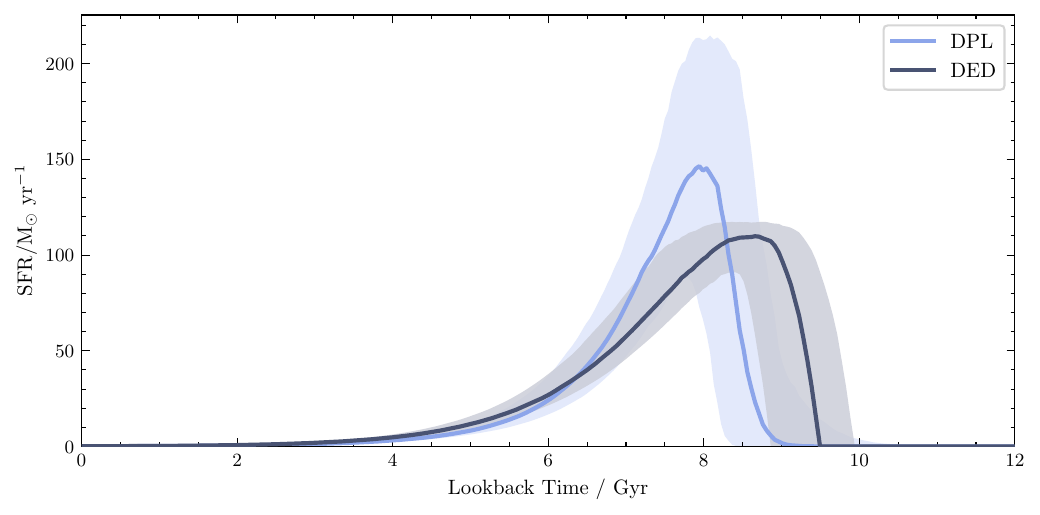}
\caption{Comparison of the \texttt{TARGETID} = 2851244993413120's reconstructed star formation history profiles between the DED (dark blue) and DPL (light blue) models. The horizontal axis represents the lookback time in Gyr, and the vertical axis shows the star formation rate (SFR) in $M_{\odot} \text{ yr}^{-1}$. In both cases, the solid lines indicate the median of the posterior distribution, while the shaded regions represent the $1\sigma$ uncertainties. 
\label{fig:SFH}}
\end{figure*}

We assess the convergence of the results and employ Hartigan \& Hartigan's dip test to evaluate the unimodality (single peak) of the posterior distributions \citep{hartigan1985dip, hartigan1985algorithm}. The dip test, with the null hypothesis that the distribution is unimodal, measures multimodality by computing the maximum difference, across all sample points, between the empirical distribution function and the unimodal distribution function that minimizes this maximum difference. A p-value threshold of 0.05 for the physical properties (age, $\tau$, formed mass, metallicity, redshift, stellar velocity dispersion ($\sigma_{\rm vel}$)) is adopted to ensure unimodality of the posterior distributions. To assess whether a posterior is significantly skewed towards the parameter space boundary, we calculate the adjusted Fisher-Pearson standardized moment coefficient of the posterior distributions. We impose a requirement that the skewness of the redshift and $\sigma_{\rm vel}$ posterior distributions be less than one, while the skewness of the remaining parameters must be below 1.5. If these requirements are not met, the results can not be considered valid, and are discarded as bad fit. A typical good fitting result is shown in Figure \ref{fig:CC} \added{and its associated corner plot in Figure \ref{fig:cornerplot}.} Two inferred SFHs are illustrated in Figure \ref{fig:SFH}.

\begin{deluxetable*}{llllll}
\tabletypesize{\footnotesize}
\tablecaption{Parameters and priors for the model we fit to our data. For Gaussian priors, $\mu$ is the mean and $\sigma$ the standard deviation of the prior distribution. Logarithmic priors are uniform in log base ten of the parameter. $z_{spec}$ is the redshift obtained from the catalog. For the metallicity prior, numerous literature illustrated that CCs have slightly super-solar metallicity \citep{2012ApJ...755...26O, 2014ApJ...780...33C, 2019ApJ...880L..31K, 2022LRR....25....6M}. Thus, we impose an 1.3 in mean and a wide $\sigma = 0.8$. \label{tab:para}}
\tablewidth{0pt}
\tablehead{
\colhead{Component} & \colhead{Parameter} & \colhead{Symbol / Unit} & \colhead{Range or Values} & \colhead{Prior} & \colhead{Hyperparameters} \\
}
\startdata
Global & Redshift & $z$ & $z_{spec} \pm 3\sigma$ & Gaussian & $\mu = z_{spec},\, \sigma = 0.01$\\
& Velocity dispersion & $\sigma_{\rm vel}\ /\ {\rm km\ s^{-1}}$ & $(10,\,400)$ & logarithmic & \\
& Ionization parameter & & $\log_{10}(U) = -3$ &  & \\
\tableline
SFH & Formed stellar mass & $M_{\rm formed} /\ M_\odot$ & $(1,\,10^{13})$ & logarithmic & \\
& Metallicity & $Z\ /\ Z_\odot$ & $(0,\,3)$ & Gaussian & $\mu=1.3,\, \sigma=0.8$ \\
DED & Age & $\rm Age\ /\ {\rm Gyr}$ & $(0.1,\,20)$ & logarithmic & \\
& Star formation timescale & $\tau\ /\ {\rm Gyr}$ & $(0,\,1)$ & uniform & \\
DPL & Falling slope & $\alpha$ & $(0.1,\,1000)$ & logarithmic & \\
& Rising slope & $\beta$ & $(10,\,1000)$ & logarithmic & \\
& Peak time & $\tau\ /\ {\rm Gyr}$ & $(0,\,20)$ & uniform & \\
\tableline
Dust & Attenuation at 5500\AA & $A_V\ /\ {\rm mag}$ & $(0,\,8)$ & uniform & \\
& Attenuation factor & $\epsilon$ & $2$ & & \\
&Lifetime of stellar birth clouds& $t_{BC}\ /\ {\rm Gyr}$ & $0.01$ & & \\
\tableline
Calibration & Zero order & $P_0$ & $(2,\,50)$ & uniform & \\
& First order & $P_1$ & $(-0.5,\,0.5)$ & Gaussian & $\mu=0,\, \sigma=0.25$ \\
& Second order & $P_2$ & $(-0.5,\,0.5)$ & Gaussian & $\mu=0,\, \sigma=0.25$ \\
\tableline
Noise & White noise scaling & $a$ & $(0.1,\,10)$ & logarithmic & \\
& Correlated noise amplitude & $b\ /\ f_{\max}$ & $(0.0001,\,1)$ & logarithmic & \\
& Correlation length & $l\ /\ \Delta\lambda$ & $(0.006,\,1)$ & logarithmic & \\
\enddata
\end{deluxetable*}

\section{Result} \label{result}

\subsection{Galaxy properties}

\begin{figure*}[ht!]
\plotone{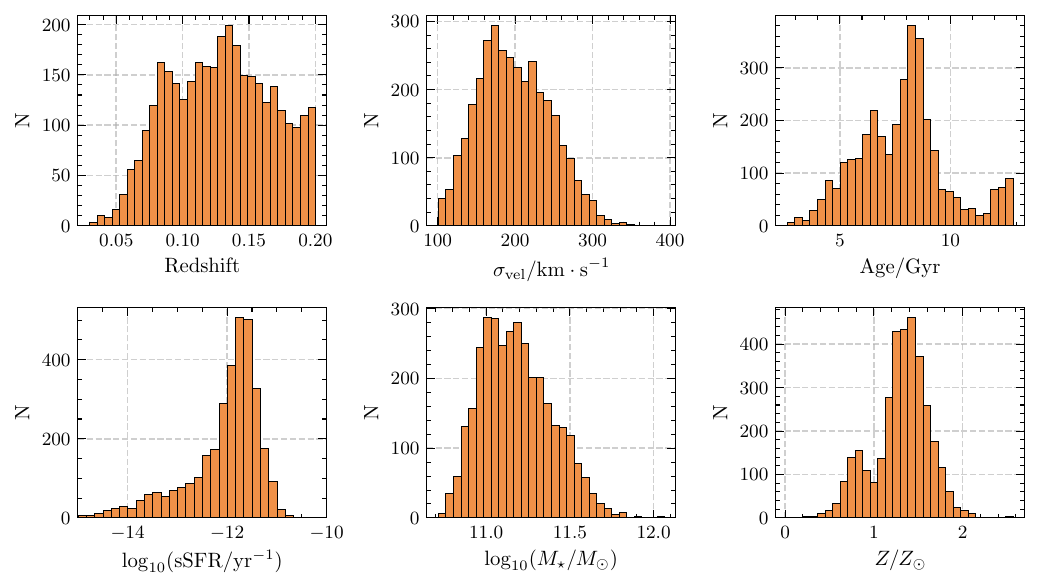}
\caption{Distributions of the inferred physical properties for our CC sample, derived from \texttt{BAGPIPES}. The panels show the distribution of posterior median values for the spectroscopic redshift, stellar velocity dispersion ($\sigma_{\rm vel}$), stellar age, sSFR ($\log_{10}(\rm sSFR/yr^{-1})$), stellar mass ($\log_{10}(M_{\star}/M_{\odot})$) and metallicity relative to solar ($Z/Z_{\odot}$), respectively.
\label{fig:property}}
\end{figure*}

\begin{figure*}[ht!]
\plotone{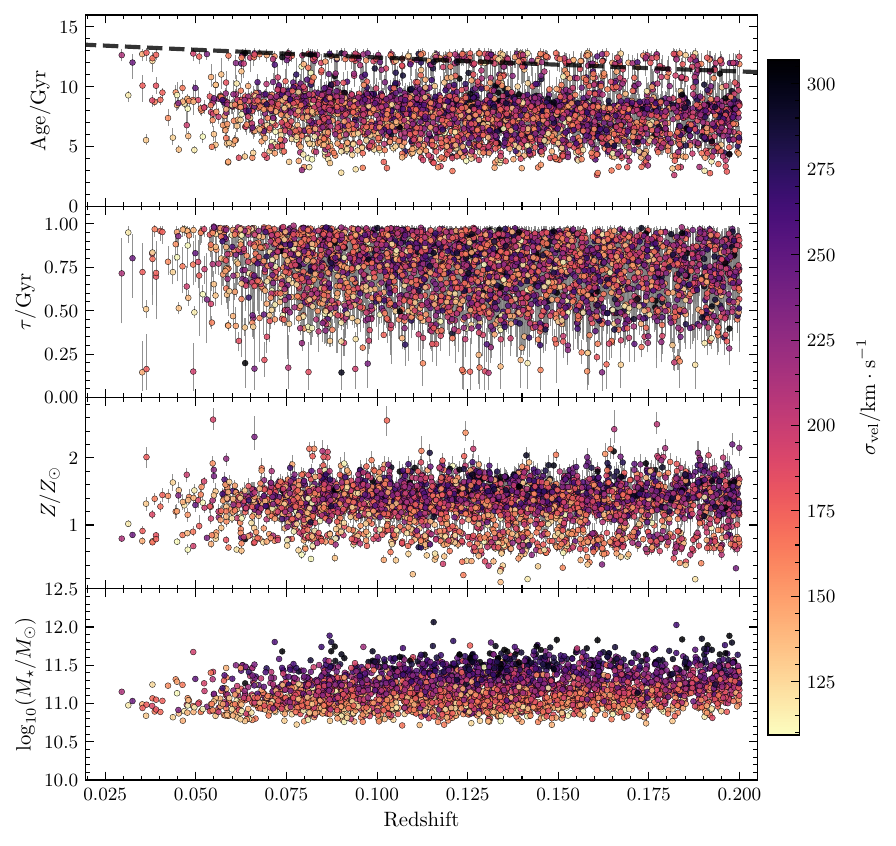}
\caption{Redshift distribution of stellar age, star formation timescale ($\tau$), metallicity and logarithmic stellar mass ($\log_{10}(M_{\star}/M_{\odot})$) obtained from the full-spectrum fitting of our CC sample. Each galaxy is color-coded by its velocity dispersion ($\sigma_{\rm vel}$). The black dashed line in the first panel is the  age--redshift relation predicted by the \cite{2020A&A...641A...6P} $\Lambda$CDM model.
\label{fig:property_fof_redsft}}
\end{figure*}

Following the full-spectrum fitting procedure, we impose a stringent mass and $\sigma_{\rm vel}$ threshold of $\log_{10}(M_{\rm formed} / M_{\odot}) > 11$ and $\sigma_{\rm vel} > 100$ km/s to ensure the homogeneity of our CC sample, which results in 3429 CCs in total\footnote{Note that formed stellar mass ($M_{\rm formed}$) is not the same as living stellar mass ($M_{\star}$) in \texttt{BAGPIPES}. $M_{\rm formed}$ is defined as $M_{\rm{formed}}=\int_{0}^{t}\rm{SFR}(t^{\prime})\mathrm{d}t^{\prime}.$, which represents the whole stellar mass formed from $t=0$ to the time $t$. $M_{\star}$ on the other hand excludes remnants.}. The distributions of the main physical parameters (redshift, $\sigma_{\rm vel}$, age, sSFR, stellar mass, metallicity) are illustrated in Figure \ref{fig:property}. Four parameters (age, $\tau$, metallicity and stellar mass) are illustrated in Figure \ref{fig:property_fof_redsft} as a function of redshift. Each galaxy is color-coded by its median $\sigma_{\rm vel}$. Unless otherwise specified, we report the physical properties using the median values along with the $16^{\rm th}$ and $84^{\rm th}$ percentiles. Overall, our CC sample is characterized by:

\begin{itemize}
    \item High stellar masses: $\log_{10}(M_{\star}/M_{\odot})=11.16^{+0.25}_{-0.19}$;
    \item Low dust attenuation and sSFR: $A_V = 0.10^{+0.13}_{-0.07}$ mag and sSFR is $\log_{10}(\rm sSFR/yr^{-1}) < -10$, which are typical for massive and passive galaxies;
    \item Rapid SFH: $\tau = 0.76^{+0.15}_{-0.23}$ Gyr;
    \item Super-solar metallicities: $Z/Z_\odot = 1.33^{+0.24}_{-0.44}$, we can clearly see a double \added{peak} on the metallicity distribution. The majority of our sample is slightly above solar, while there is a tiny peak below the solar metallicity. This secondary, metal-poor peak likely \added{originates} from the dust--metallicity degeneracy in fitting \citep{2023ApJS..265...48J} or may suggest the presence of a sub-population within our CC sample which has slightly sub-solar metallicity \citep{2019ApJ...880L..31K, 2022ApJ...929..131C}, likely reflecting different chemical enrichment histories. 
\end{itemize}

Remarkably, even without invoking a cosmological prior, the derived stellar ages for the majority of our sample exhibit a declining trend with redshift that is consistent with the $\Lambda$CDM model \citep{2020A&A...641A...6P}. We identify a clear positive correlation between stellar mass and $\sigma_{\rm vel}$. Regarding the SFH timescale $\tau$, we find no significant evolution with redshift; however, our sample recovers the ``mass-downsizing'' scenario, where more massive galaxies characterize shorter formation timescales \citep{1996AJ....112..839C, 2010MNRAS.404.1775T, 2017MNRAS.469.3108C}. Besides, the lack of significant redshift dependence in metallicity and mass further validates our selection criteria in isolating a synchronized population of CCs.

In summary, we have constructed a robust and representative CC sample. In the following section, we will discuss the computation of $H(z)$. 

\subsection{Computing H(z) measurement}

\begin{figure*}[ht!]
\plotone{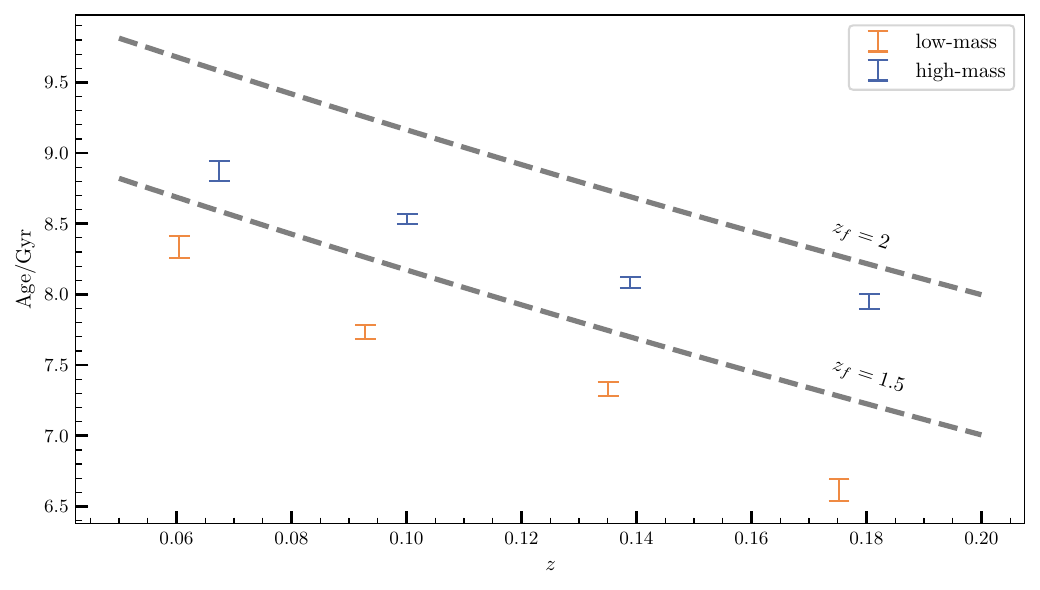}
\caption{Median-binned age--redshift relations for our baseline results. The blue and orange points represent the lower and higher mass bins divided by the median value. For illustration, we also present the redshift $z_f$ of galaxy formation (gray dashed lines) within the $\Lambda$CDM model by \citet{2020A&A...641A...6P}.
\label{fig:age-redshift}}
\end{figure*}

To derive $H(z)$ from Equation \ref{H(z)}, first, we need to construct the median age--redshift relation using the selected CC sample. The sample is partitioned into both redshift and stellar mass intervals. Within each sub-sample, we determine the median age and the corresponding mean redshift. The uncertainty associated with the age is estimated as ${\rm MAD} / \sqrt{n}$, where MAD denotes the median absolute deviation and $n$ is the number of galaxies per bin. To mitigate progenitor bias \citep{1995IAUS..164..269F, 2000ApJ...541...95V}, we divide the sample into two mass bins according to the median value, $\log_{10}(M_{\star}/M_{\odot})=11.16$. Here we choose to split the redshift into four equally spaced intervals and we consider this scheme as the benchmark for our research. The age--redshift relation for the benchmark case is presented in Figure \ref{fig:age-redshift} and the values are reported in Table \ref{tab:age-redshift}.

\begin{deluxetable*}{cccc}
\tabletypesize{\footnotesize}
\tablecaption{Median ages and properties of the selected CC sample considering the benchmark binning. Mean values of redshift, median ages and the number of galaxies in each bin (n) are reported. \label{tab:age-redshift}}
\tablewidth{0pt}
\tablehead{
&\colhead{Redshift} & \colhead{Age/Gyr} & \colhead{n} \\
}
\startdata
         & 0.06 & $8.33\pm 0.08$ & 178\\
       low-mass  & 0.09 & $7.73\pm 0.05$ & 590\\
         & 0.14 & $7.33\pm 0.05$& 617\\
         & 0.17 & $6.62\pm 0.07$& 330\\
         \tableline
         & 0.07 & $8.87\pm0.07$ & 106\\
       high-mass  & 0.1 & $8.54\pm0.04$ & 479\\
         & 0.14 & $8.08\pm0.04$ & 615\\
         & 0.18 & $7.95\pm0.05$ & 514\\
\enddata
\end{deluxetable*}

The next step is to individually calculate the $H(z)$ points on every mass bin. When estimating the redshift--age derivative ($\frac{dz}{dt}$), we adopt a non-adjacent differencing approach to minimize potential covariance rather than using consecutive bins. Specifically, for a total of $N$ redshift bins within a given mass range, we compute the difference between the $i^{\rm th}$ and the $(i + N/2)^{\rm th}$ data points. This strategy necessitates an even number of redshift bins, ensuring that each data point is utilized uniquely to avoid spurious correlations between the derived $H(z)$ values \citep{2023ApJS..265...48J}. 

Finally, these values are inverse variance averaged to get a single and more accurate estimate of $H(z)$\citep{2016JCAP...05..014M, 2022ApJ...928L...4B, 2023A&A...679A..96T, 2023ApJS..265...48J}. In this way, our $H(z)$ estimation of our benchmark binning is 
\[H(z = 0.12) = 71.33 \pm 3.45(\rm stat.) \, km/s/Mpc.\]
The associated error is only statistical and is rather low due to our large CC sample size. In the next section, different binning strategies and SFHs are compared in order to access the systematic effect on our $H(z)$ measurement. 

\subsection{Discussion on systematic effects}
The major sources of systematic uncertainty are binning method, SFH choice and the SPS model \citep{2022LRR....25....6M, 2023A&A...679A..96T, 2023ApJS..265...48J}. In this research, we did not consider the SPS model effect. We evaluate the impact of different binning configurations by testing four distinct strategies:
\begin{itemize}
\item Binning type: either equally spaced (fixed) or equally populated (quant.);
\item Binning size: choosing four intervals or two intervals of redshift.
\end{itemize}
The configuration utilizing equally spaced and dividing into four bins is designated as the benchmark, while the remaining three strategies serve as comparative cases to assess systematic effects. Other measurements are shown in Table \ref{tab:sys}. Furthermore, results with DPL are also listed to access the SFH contribution of systematics. 

\begin{deluxetable*}{lllc}
\tabletypesize{\footnotesize}
\tablecaption{Measurements of $H(z)$ obtained applying the CC
method with different binning and SFH configurations.
\label{tab:sys}}
\tablewidth{0pt}
\tablehead{
\colhead{SFH}&\colhead{Binning} & \colhead{$z$} & \colhead{$H(z)/\rm km\cdot s^{-1}Mpc^{-1}$} \\
}
\startdata
    DED & 4 fixed & 0.12 & $71.33\pm3.45$\\
    \tableline
        & 4 quant. & 0.13 & $70.01\pm3.86$\\
        & 2 fixed & 0.12 & $69.42\pm3.71$\\
        & 2 quant. & 0.13 & $69.44\pm4.08$\\
    \tableline
    DPL & 4 fixed & 0.12 & $73.00\pm4.16$\\
\enddata
\end{deluxetable*}

The discrepancy in different binning strategies is included as the mean value of difference subtracted by benchmark, $\Delta H_{\rm bin}=\pm1.71\, \rm km\cdot s^{-1}Mpc^{-1}$. The SFH contribution to systematic is $\Delta H_{\rm SFH}=\pm1.67\, \rm km\cdot s^{-1}Mpc^{-1}$. Further discussion on the SFH effect is shown in the Appendix \ref{apx: A}. 

\begin{figure*}[ht!]
\plotone{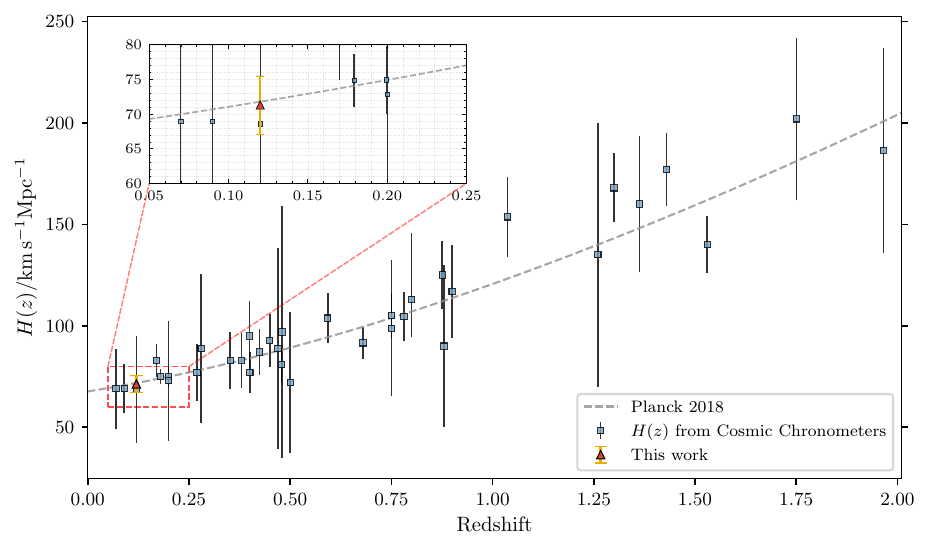}
\caption{Final $H(z)$ measurement of this work in comparison with all the $H(z)$ estimations obtained up to now with the cosmic chronometer method \citep{2005PhRvD..71l3001S, 2010JCAP...02..008S, Zhang_2014, 2012JCAP...08..006M, 2015MNRAS.450L..16M, 2016JCAP...05..014M, 2017MNRAS.467.3239R, 2022ApJ...928L...4B, 2023ApJS..265...48J, 2023A&A...679A..96T, 2023JCAP...11..047J, 2025MNRAS.540.3135L}. The gray dashed line represents the $H(z)$ prediction by the $\Lambda$CDM model \citep{2020A&A...641A...6P}.
\label{fig:Hz}}
\end{figure*}

Finally, summing the errors in quadrature, our measurement at redshift 0.12 is 
\[H(z = 0.12) = 71.33 \pm 4.20\, \rm km\cdot s^{-1}Mpc^{-1}\]
This is also shown with other $H(z)$ measurements in Figure \ref{fig:Hz}. 

\section{conclusion}

In this paper, we estimate $H(z)$ at redshift 0.12 with the cosmic chronometers selected from DESI DR1. We derive their physical properties from a joint fitting analysis with spectra and photometry using a modified version of public code \texttt{BAGPIPES} which removed its implemented cosmological prior \citep{2018MNRAS.480.4379C, 2023ApJS..265...48J}. Our main results are summarized as follows. 

\begin{enumerate}
    \item We first select a sample of massive and passively evolving galaxies with the Stellar Mass and Emission Line Catalog combining multiple criteria (galaxy morphology, specific star formation rate, stellar mass, $H/K$ indices, emission line);
    \item We then perform a joint fitting with two types of SFHs and find that age of most our sample are decreasing with redshift in agreement with the $\Lambda$CDM cosmological model. At fixed redshift, more massive galaxies result older than lower mass ones, confirming the mass-downsizing trend at low redshift.
    \item With the robust cosmic chronometer sample, we construct an precise age--redshift relation and obtain a new measurement of \(H(z = 0.12) = 71.33\pm3.45(\rm stat.) \, km/s/Mpc\). We also test that our result is consistent with other literature, as well as with the prediction of a $\Lambda$CDM model assuming
    \item We assess the systematic uncertainty of our measurement by varying the binning strategies and SFH in the fit. We estimate a systematic error of $\Delta H_{\rm sys}=\rm \pm 1.71(bin) \pm 1.67 (SFH)\,km/s/Mpc$, mainly dominated by different binning plans.
    \item In the end, we obtain a measurement of the Hubble parameter  at redshift 0.12, $H(z = 0.12) = 71.33 \pm 4.20\, \rm km/s/Mpc.$   
\end{enumerate}

To conclude, this paper consolidate the CC method using full-spectrum fitting technique. Further investigation can be focused on different SPS model \citep{2009ApJ...699..486C, 2010ApJ...712..833C} and upcoming survey like Euclid \citep{2011arXiv1110.3193L} and Wide-field Spectroscopic Telescope \citep{2024arXiv240512518B, 2025arXiv251222964T} that will significantly improve the census of massive and passive galaxies, especially at $z >$ 1. 

\begin{acknowledgments}
We thank Dr.Kang Jiao in Zhengzhou University, China and Dr.Adam Carnall in the Institute for Astronomy at Edinburgh University for their useful and constructive feedback to this work. Also, the authors acknowledge \href{https://paratera.com/}{Beijng PARATERA Tech CO.,Ltd.} for providing HPC resources that have contributed to our research. This work is supported by the National Natural Science Foundation of China (No.12233011).

This research uses services or data provided by the SPectra Analysis and Retrievable Catalog Lab (SPARCL) and the Astro Data Lab, which are both part of the Community Science and Data Center (CSDC) program at NSF National Optical-Infrared Astronomy Research Laboratory. NOIRLab is operated by the Association of Universities for Research in Astronomy (AURA), Inc. under a cooperative agreement with the National Science Foundation.

This research used data obtained with the Dark Energy Spectroscopic Instrument (DESI). DESI construction and operations is managed by the Lawrence Berkeley National Laboratory. This material is based upon work supported by the U.S. Department of Energy, Office of Science, Office of High-Energy Physics, under Contract No. DE--AC02--05CH11231, and by the National Energy Research Scientific Computing Center, a DOE Office of Science User Facility under the same contract. Additional support for DESI was provided by the U.S. National Science Foundation (NSF), Division of Astronomical Sciences under Contract No. AST--0950945 to the NSF's National Optical-Infrared Astronomy Research Laboratory; the Science and Technology Facilities Council of the United Kingdom; the Gordon and Betty Moore Foundation; the Heising--Simons Foundation; the French Alternative Energies and Atomic Energy Commission (CEA); the National Council of Humanities, Science and Technology of Mexico (CONAHCYT); the Ministry of Science and Innovation of Spain (MICINN), and by the DESI Member Institutions: \url{https://www.desi.lbl.gov/collaborating-institutions/}. The DESI collaboration is honored to be permitted to conduct scientific research on I'oligam Du'ag (Kitt Peak), a mountain with particular significance to the Tohono O'odham Nation. Any opinions, findings, and conclusions or recommendations expressed in this material are those of the author(s) and do not necessarily reflect the views of the U.S. National Science Foundation, the U.S. Department of Energy, or any of the listed funding agencies. 

\end{acknowledgments}

%
\facility{Astro Data Lab, DESI}

\software{SPARCL (Juneau et al. 2024), BAGPIPES \citep{2018MNRAS.480.4379C}, NumPy \citep{harris2020array}, Astropy \citep{astropy:2013, astropy:2018, astropy:2022}, SciPy \citep{2020SciPy-NMeth}, pandas \citep{mckinney-proc-scipy-2010}, pyLick \citep{2022ApJ...927..164B}}


\appendix

\section{The DPL SFH results} \label{apx: A}

\begin{figure*}[ht!]
\plotone{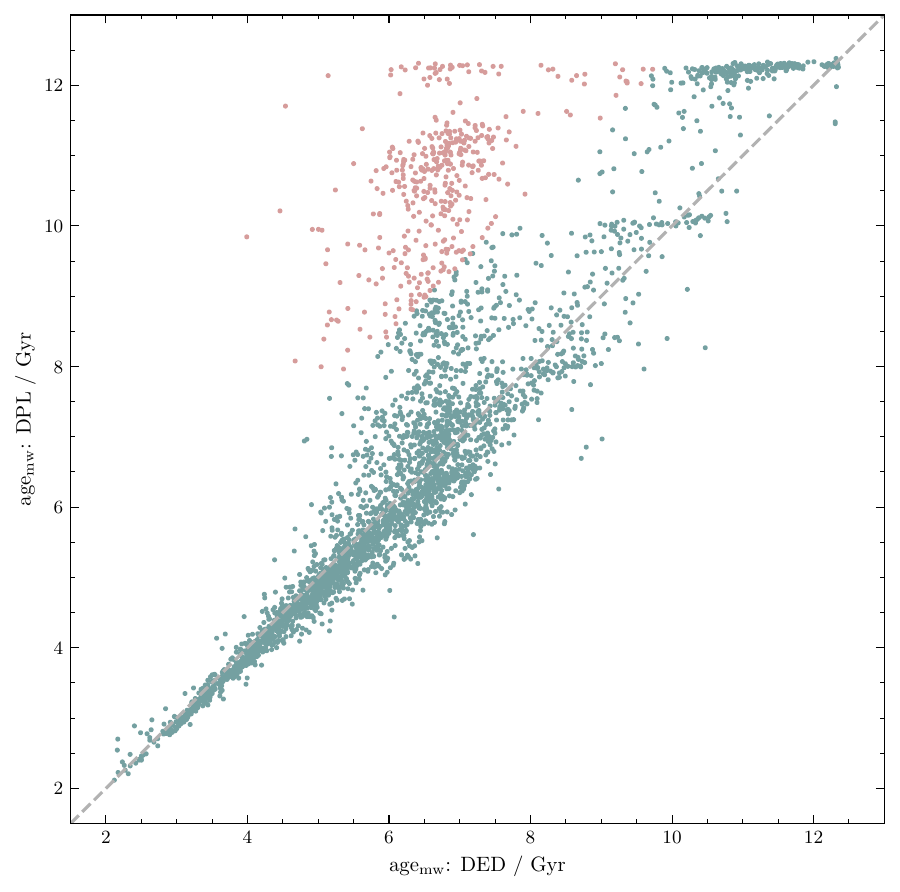}
\caption{The comparison of mass-weighted ages in two SFHs. The light-red points are excluded as outliers. The gray dashed line is the one-to-one relation. 
\label{fig:age-age}}
\end{figure*}

Even though we have imposed a conservative prior (See Table \ref{tab:para} and Section \ref{subsubsec: SFH}) in the DPL SFH configuration, it is still inevitable to arouse the same issue appeared in the \cite{2023A&A...679A..96T}. In this context, we suppose the major reason is the lack of photometry observations compared to other research (e.g. \cite{2023A&A...679A..96T, 2023ApJS..265...48J}). We make comparison of mass--weighted age in two SFH choice in Figure \ref{fig:age-age} and exclude 3$\sigma$ outliers to build up our CC sample in DPL measurement. 

\section{Discussion on the noise model} \label{apx: B}

\begin{figure*}[ht!]
\plotone{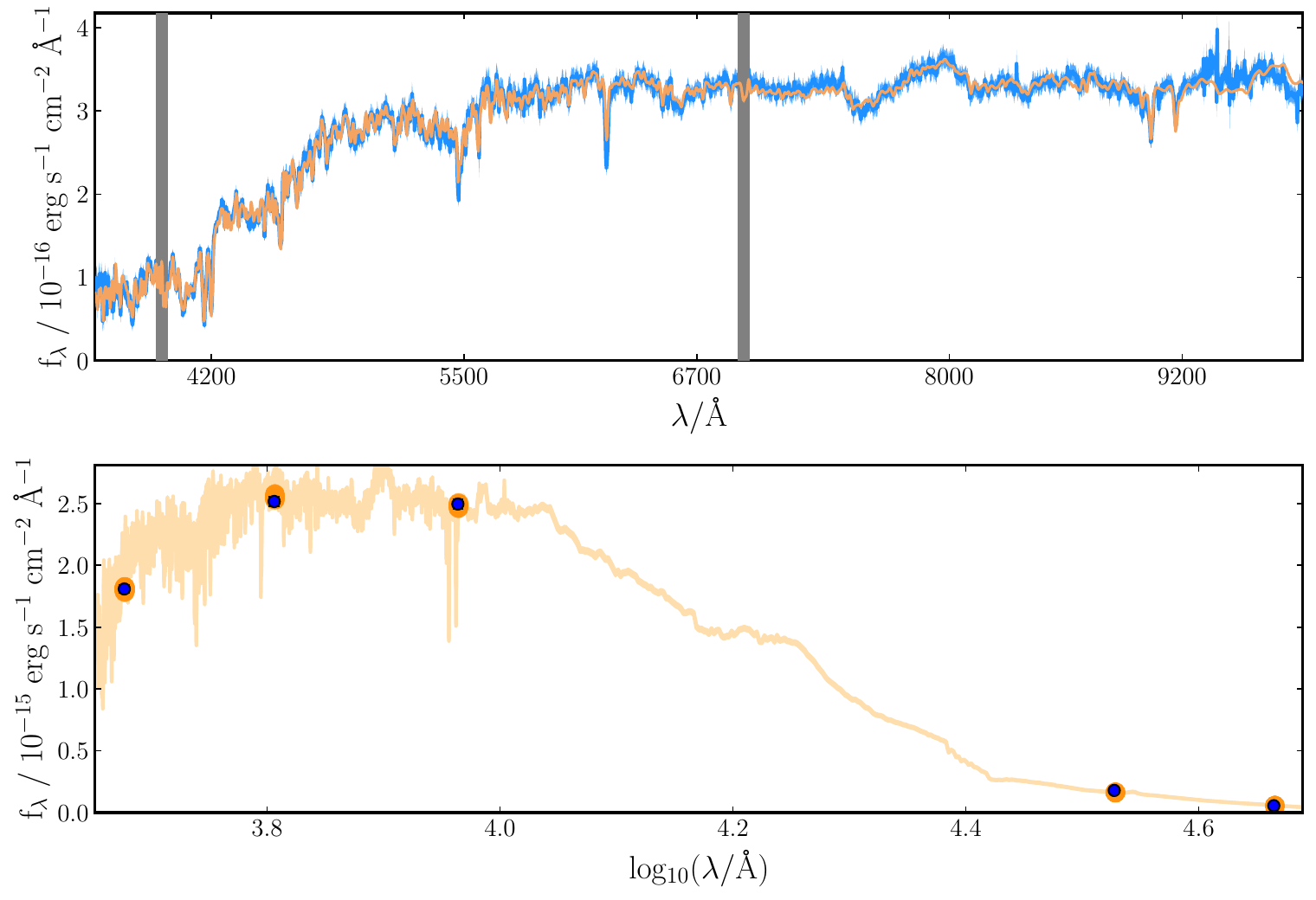}
\caption{\texttt{TARGETID} = 2851244993413120 fitting result only with white noise.
\label{fig:white noise}}
\end{figure*}

\added{During the full-spectrum fitting procedure in section \ref{fitting}, we deploy a Gaussian process (GP) noise model. The median of hyperparameter $l$ of our sample is 0.011 and the median of $b$ is 0.014 as shown in the figure \ref{fig:noise} below. The major reason we choose this model is that GP model behaves better in red side of the spectrum compared to the white noise (see Figure \ref{fig:CC} v.s. Figure \ref{fig:white noise}). However, its substantial computational expense remains a significant drawback. To address this, the latest version of \texttt{BAGPIPES} implements a more efficient algorithm proposed by \cite{2024MNRAS.528.4029L}.}

\begin{figure*}[ht!]
\plotone{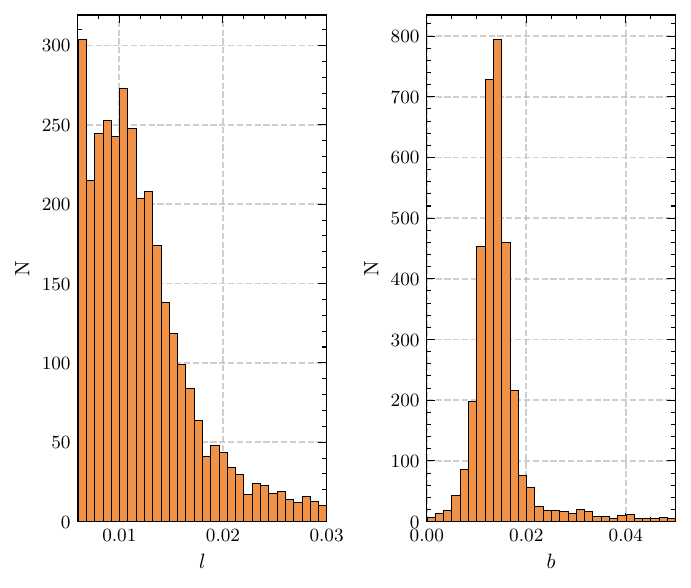}
\caption{Distributions of the inferred hyperparameters for our CC sample. The panels show the distribution of posterior median values for $l$ and $b$, respectively.
\label{fig:noise}}
\end{figure*}

\section{posterior distribution corner plot of a typical CC}

\added{The figure \ref{fig:cornerplot} below shows the posterior distribution corner plot of all fitted parameters.}

\begin{figure*}[ht!]
\plotone{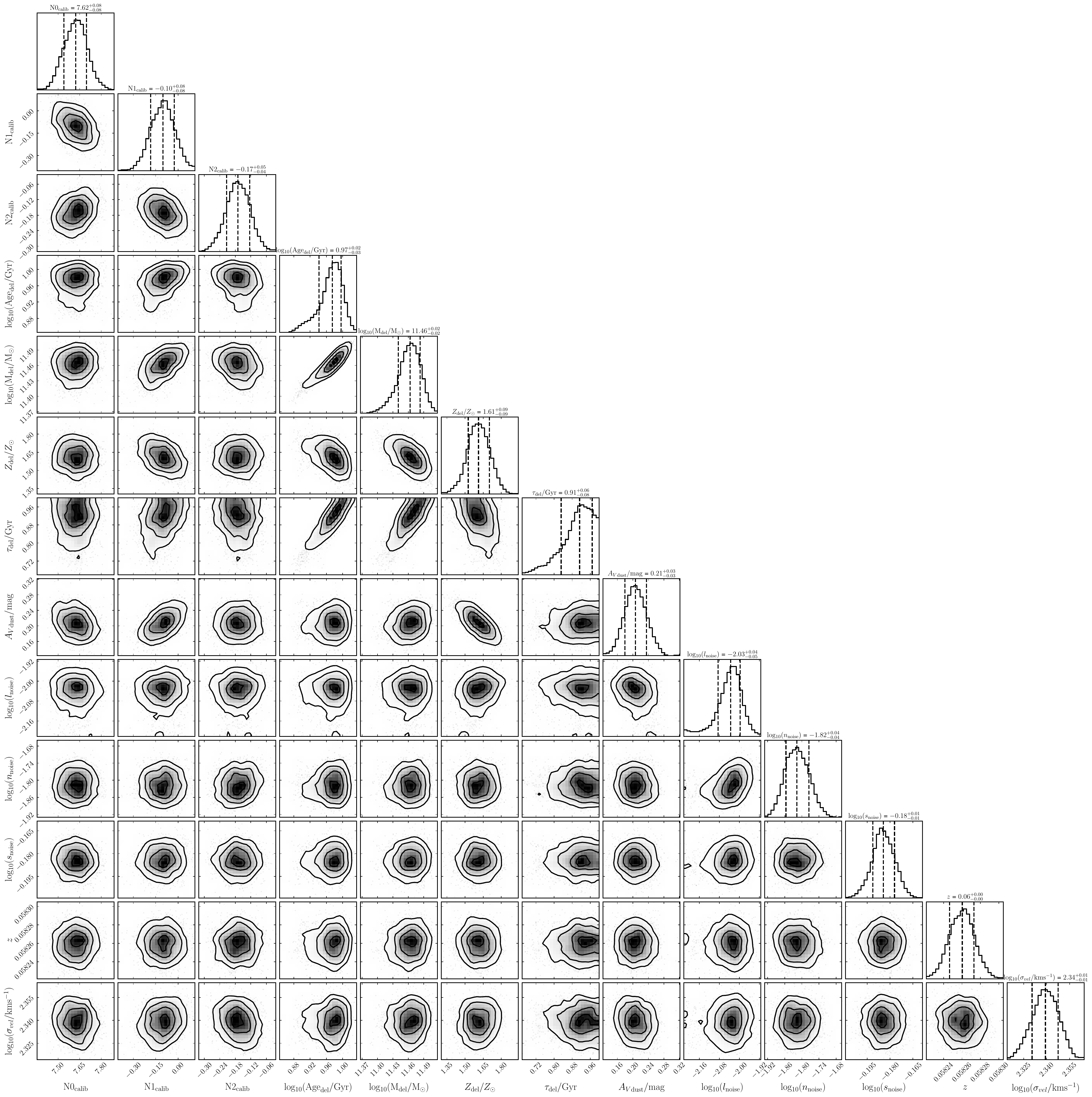}
\caption{\added{The posterior distribution corner plot of all fitted parameters (\texttt{TARGETID} = 2851244993413120).}
\label{fig:cornerplot}}
\end{figure*}


\bibliography{sample701}{}

@ARTICLE{2024MNRAS.528.4029L,
       author = {{Leung}, Ho-Hin and {Wild}, Vivienne and {Papathomas}, Michail and {Carnall}, Adam and {Zheng}, Yirui and {Boardman}, Nicholas and {Wang}, Cara and {Johansson}, Peter H.},
        title = "{Chemical evolution of local post-starburst galaxies: implications for the mass-metallicity relation}",
      journal = {\mnras},
         year = 2024,
        month = mar,
       volume = {528},
       number = {3},
        pages = {4029-4052},
          doi = {10.1093/mnras/stae225},
archivePrefix = {arXiv},
       eprint = {2309.16626},
 primaryClass = {astro-ph.GA},
       adsurl = {https://ui.adsabs.harvard.edu/abs/2024MNRAS.528.4029L}
}

@ARTICLE{2025arXiv250819081W,
       author = {{Wang}, Yi-Ying and {Lei}, Lei and {Tang}, Shao-Peng and {Fan}, Yi-Zhong},
        title = "{Lensing amplitude anomaly and varying electron mass alleviate the Hubble and $S_8$ tensions}",
      journal = {\jcap},
         year = 2026,
        volume = {2026},
        number = {01},
        pages = {009},
          doi = {10.1088/1475-7516/2026/01/009},
archivePrefix = {arXiv},
       eprint = {2508.19081},
 primaryClass = {astro-ph.CO},
    url = {https://doi.org/10.1088/1475-7516/2026/01/009},
       adsurl = {https://ui.adsabs.harvard.edu/abs/2025arXiv250819081W}
}

@ARTICLE{1992ApJ...396L...1S,
       author = {{Smoot}, G.~F. and {Bennett}, C.~L. and {Kogut}, A. and {Wright}, E.~L. and {Aymon}, J. and {Boggess}, N.~W. and {Cheng}, E.~S. and {de Amici}, G. and {Gulkis}, S. and {Hauser}, M.~G. and {Hinshaw}, G. and {Jackson}, P.~D. and {Janssen}, M. and {Kaita}, E. and {Kelsall}, T. and {Keegstra}, P. and {Lineweaver}, C. and {Loewenstein}, K. and {Lubin}, P. and {Mather}, J. and {Meyer}, S.~S. and {Moseley}, S.~H. and {Murdock}, T. and {Rokke}, L. and {Silverberg}, R.~F. and {Tenorio}, L. and {Weiss}, R. and {Wilkinson}, D.~T.},
        title = "{Structure in the COBE Differential Microwave Radiometer First-Year Maps}",
      journal = {\apjl},
         year = 1992,
        month = sep,
       volume = {396},
        pages = {L1},
          doi = {10.1086/186504},
       adsurl = {https://ui.adsabs.harvard.edu/abs/1992ApJ...396L...1S}
}

@ARTICLE{2020A&A...641A...6P,
       author = {{Planck Collaboration} and {Aghanim}, N. and {Akrami}, Y. and {Ashdown}, M. and {Aumont}, J. and {Baccigalupi}, C. and {Ballardini}, M. and {Banday}, A.~J. and {Barreiro}, R.~B. and {Bartolo}, N. and {Basak}, S. and {Battye}, R. and {Benabed}, K. and {Bernard}, J.-P. and {Bersanelli}, M. and {Bielewicz}, P. and {Bock}, J.~J. and {Bond}, J.~R. and {Borrill}, J. and {Bouchet}, F.~R. and {Boulanger}, F. and {Bucher}, M. and {Burigana}, C. and {Butler}, R.~C. and {Calabrese}, E. and {Cardoso}, J.-F. and {Carron}, J. and {Challinor}, A. and {Chiang}, H.~C. and {Chluba}, J. and {Colombo}, L.~P.~L. and {Combet}, C. and {Contreras}, D. and {Crill}, B.~P. and {Cuttaia}, F. and {de Bernardis}, P. and {de Zotti}, G. and {Delabrouille}, J. and {Delouis}, J.-M. and {Di Valentino}, E. and {Diego}, J.~M. and {Dor{\'e}}, O. and {Douspis}, M. and {Ducout}, A. and {Dupac}, X. and {Dusini}, S. and {Efstathiou}, G. and {Elsner}, F. and {En{\ss}lin}, T.~A. and {Eriksen}, H.~K. and {Fantaye}, Y. and {Farhang}, M. and {Fergusson}, J. and {Fernandez-Cobos}, R. and {Finelli}, F. and {Forastieri}, F. and {Frailis}, M. and {Fraisse}, A.~A. and {Franceschi}, E. and {Frolov}, A. and {Galeotta}, S. and {Galli}, S. and {Ganga}, K. and {G{\'e}nova-Santos}, R.~T. and {Gerbino}, M. and {Ghosh}, T. and {Gonz{\'a}lez-Nuevo}, J. and {G{\'o}rski}, K.~M. and {Gratton}, S. and {Gruppuso}, A. and {Gudmundsson}, J.~E. and {Hamann}, J. and {Handley}, W. and {Hansen}, F.~K. and {Herranz}, D. and {Hildebrandt}, S.~R. and {Hivon}, E. and {Huang}, Z. and {Jaffe}, A.~H. and {Jones}, W.~C. and {Karakci}, A. and {Keih{\"a}nen}, E. and {Keskitalo}, R. and {Kiiveri}, K. and {Kim}, J. and {Kisner}, T.~S. and {Knox}, L. and {Krachmalnicoff}, N. and {Kunz}, M. and {Kurki-Suonio}, H. and {Lagache}, G. and {Lamarre}, J.-M. and {Lasenby}, A. and {Lattanzi}, M. and {Lawrence}, C.~R. and {Le Jeune}, M. and {Lemos}, P. and {Lesgourgues}, J. and {Levrier}, F. and {Lewis}, A. and {Liguori}, M. and {Lilje}, P.~B. and {Lilley}, M. and {Lindholm}, V. and {L{\'o}pez-Caniego}, M. and {Lubin}, P.~M. and {Ma}, Y.-Z. and {Mac{\'\i}as-P{\'e}rez}, J.~F. and {Maggio}, G. and {Maino}, D. and {Mandolesi}, N. and {Mangilli}, A. and {Marcos-Caballero}, A. and {Maris}, M. and {Martin}, P.~G. and {Martinelli}, M. and {Mart{\'\i}nez-Gonz{\'a}lez}, E. and {Matarrese}, S. and {Mauri}, N. and {McEwen}, J.~D. and {Meinhold}, P.~R. and {Melchiorri}, A. and {Mennella}, A. and {Migliaccio}, M. and {Millea}, M. and {Mitra}, S. and {Miville-Desch{\^e}nes}, M.-A. and {Molinari}, D. and {Montier}, L. and {Morgante}, G. and {Moss}, A. and {Natoli}, P. and {N{\o}rgaard-Nielsen}, H.~U. and {Pagano}, L. and {Paoletti}, D. and {Partridge}, B. and {Patanchon}, G. and {Peiris}, H.~V. and {Perrotta}, F. and {Pettorino}, V. and {Piacentini}, F. and {Polastri}, L. and {Polenta}, G. and {Puget}, J.-L. and {Rachen}, J.~P. and {Reinecke}, M. and {Remazeilles}, M. and {Renzi}, A. and {Rocha}, G. and {Rosset}, C. and {Roudier}, G. and {Rubi{\~n}o-Mart{\'\i}n}, J.~A. and {Ruiz-Granados}, B. and {Salvati}, L. and {Sandri}, M. and {Savelainen}, M. and {Scott}, D. and {Shellard}, E.~P.~S. and {Sirignano}, C. and {Sirri}, G. and {Spencer}, L.~D. and {Sunyaev}, R. and {Suur-Uski}, A.-S. and {Tauber}, J.~A. and {Tavagnacco}, D. and {Tenti}, M. and {Toffolatti}, L. and {Tomasi}, M. and {Trombetti}, T. and {Valenziano}, L. and {Valiviita}, J. and {Van Tent}, B. and {Vibert}, L. and {Vielva}, P. and {Villa}, F. and {Vittorio}, N. and {Wandelt}, B.~D. and {Wehus}, I.~K. and {White}, M. and {White}, S.~D.~M. and {Zacchei}, A. and {Zonca}, A.},
        title = "{Planck 2018 results. VI. Cosmological parameters}",
      journal = {\aap},
         year = 2020,
        month = sep,
       volume = {641},
          eid = {A6},
        pages = {A6},
          doi = {10.1051/0004-6361/201833910},
archivePrefix = {arXiv},
       eprint = {1807.06209},
 primaryClass = {astro-ph.CO},
       adsurl = {https://ui.adsabs.harvard.edu/abs/2020A&A...641A...6P}
}

@ARTICLE{1998AJ....116.1009R,
       author = {{Riess}, Adam G. and {Filippenko}, Alexei V. and {Challis}, Peter and {Clocchiatti}, Alejandro and {Diercks}, Alan and {Garnavich}, Peter M. and {Gilliland}, Ron L. and {Hogan}, Craig J. and {Jha}, Saurabh and {Kirshner}, Robert P. and {Leibundgut}, B. and {Phillips}, M.~M. and {Reiss}, David and {Schmidt}, Brian P. and {Schommer}, Robert A. and {Smith}, R. Chris and {Spyromilio}, J. and {Stubbs}, Christopher and {Suntzeff}, Nicholas B. and {Tonry}, John},
        title = "{Observational Evidence from Supernovae for an Accelerating Universe and a Cosmological Constant}",
      journal = {\aj},
         year = 1998,
        month = sep,
       volume = {116},
       number = {3},
        pages = {1009-1038},
          doi = {10.1086/300499},
archivePrefix = {arXiv},
       eprint = {astro-ph/9805201},
 primaryClass = {astro-ph},
       adsurl = {https://ui.adsabs.harvard.edu/abs/1998AJ....116.1009R}
}

@ARTICLE{2005ApJ...633..560E,
       author = {{Eisenstein}, Daniel J. and {Zehavi}, Idit and {Hogg}, David W. and {Scoccimarro}, Roman and {Blanton}, Michael R. and {Nichol}, Robert C. and {Scranton}, Ryan and {Seo}, Hee-Jong and {Tegmark}, Max and {Zheng}, Zheng and {Anderson}, Scott F. and {Annis}, Jim and {Bahcall}, Neta and {Brinkmann}, Jon and {Burles}, Scott and {Castander}, Francisco J. and {Connolly}, Andrew and {Csabai}, Istvan and {Doi}, Mamoru and {Fukugita}, Masataka and {Frieman}, Joshua A. and {Glazebrook}, Karl and {Gunn}, James E. and {Hendry}, John S. and {Hennessy}, Gregory and {Ivezi{\'c}}, Zeljko and {Kent}, Stephen and {Knapp}, Gillian R. and {Lin}, Huan and {Loh}, Yeong-Shang and {Lupton}, Robert H. and {Margon}, Bruce and {McKay}, Timothy A. and {Meiksin}, Avery and {Munn}, Jeffery A. and {Pope}, Adrian and {Richmond}, Michael W. and {Schlegel}, David and {Schneider}, Donald P. and {Shimasaku}, Kazuhiro and {Stoughton}, Christopher and {Strauss}, Michael A. and {SubbaRao}, Mark and {Szalay}, Alexander S. and {Szapudi}, Istv{\'a}n and {Tucker}, Douglas L. and {Yanny}, Brian and {York}, Donald G.},
        title = "{Detection of the Baryon Acoustic Peak in the Large-Scale Correlation Function of SDSS Luminous Red Galaxies}",
      journal = {\apj},
         year = 2005,
        month = nov,
       volume = {633},
       number = {2},
        pages = {560-574},
          doi = {10.1086/466512},
archivePrefix = {arXiv},
       eprint = {astro-ph/0501171},
 primaryClass = {astro-ph},
       adsurl = {https://ui.adsabs.harvard.edu/abs/2005ApJ...633..560E}
}

@ARTICLE{2001MNRAS.327.1297P,
       author = {{Percival}, Will J. and {Baugh}, Carlton M. and {Bland-Hawthorn}, Joss and {Bridges}, Terry and {Cannon}, Russell and {Cole}, Shaun and {Colless}, Matthew and {Collins}, Chris and {Couch}, Warrick and {Dalton}, Gavin and {De Propris}, Roberto and {Driver}, Simon P. and {Efstathiou}, George and {Ellis}, Richard S. and {Frenk}, Carlos S. and {Glazebrook}, Karl and {Jackson}, Carole and {Lahav}, Ofer and {Lewis}, Ian and {Lumsden}, Stuart and {Maddox}, Steve and {Moody}, Stephen and {Norberg}, Peder and {Peacock}, John A. and {Peterson}, Bruce A. and {Sutherland}, Will and {Taylor}, Keith},
        title = "{The 2dF Galaxy Redshift Survey: the power spectrum and the matter content of the Universe}",
      journal = {\mnras},
         year = 2001,
        month = nov,
       volume = {327},
       number = {4},
        pages = {1297-1306},
          doi = {10.1046/j.1365-8711.2001.04827.x},
archivePrefix = {arXiv},
       eprint = {astro-ph/0105252},
 primaryClass = {astro-ph},
       adsurl = {https://ui.adsabs.harvard.edu/abs/2001MNRAS.327.1297P}
}

@ARTICLE{2022ApJ...938..110B,
       author = {{Brout}, Dillon and {Scolnic}, Dan and {Popovic}, Brodie and {Riess}, Adam G. and {Carr}, Anthony and {Zuntz}, Joe and {Kessler}, Rick and {Davis}, Tamara M. and {Hinton}, Samuel and {Jones}, David and {Kenworthy}, W. D'Arcy and {Peterson}, Erik R. and {Said}, Khaled and {Taylor}, Georgie and {Ali}, Noor and {Armstrong}, Patrick and {Charvu}, Pranav and {Dwomoh}, Arianna and {Meldorf}, Cole and {Palmese}, Antonella and {Qu}, Helen and {Rose}, Benjamin M. and {Sanchez}, Bruno and {Stubbs}, Christopher W. and {Vincenzi}, Maria and {Wood}, Charlotte M. and {Brown}, Peter J. and {Chen}, Rebecca and {Chambers}, Ken and {Coulter}, David A. and {Dai}, Mi and {Dimitriadis}, Georgios and {Filippenko}, Alexei V. and {Foley}, Ryan J. and {Jha}, Saurabh W. and {Kelsey}, Lisa and {Kirshner}, Robert P. and {M{\"o}ller}, Anais and {Muir}, Jessie and {Nadathur}, Seshadri and {Pan}, Yen-Chen and {Rest}, Armin and {Rojas-Bravo}, Cesar and {Sako}, Masao and {Siebert}, Matthew R. and {Smith}, Mat and {Stahl}, Benjamin E. and {Wiseman}, Phil},
        title = "{The Pantheon+ Analysis: Cosmological Constraints}",
      journal = {\apj},
         year = 2022,
        month = oct,
       volume = {938},
       number = {2},
          eid = {110},
        pages = {110},
          doi = {10.3847/1538-4357/ac8e04},
archivePrefix = {arXiv},
       eprint = {2202.04077},
 primaryClass = {astro-ph.CO},
       adsurl = {https://ui.adsabs.harvard.edu/abs/2022ApJ...938..110B}
}

@ARTICLE{2025PhRvD.112h3515A,
       author = {{Abdul Karim}, M. and {Aguilar}, J. and {Ahlen}, S. and {Alam}, S. and {Allen}, L. and {Prieto}, C. Allende and {Alves}, O. and {Anand}, A. and {Andrade}, U. and {Armengaud}, E. and {Aviles}, A. and {Bailey}, S. and {Baltay}, C. and {Bansal}, P. and {Bault}, A. and {Behera}, J. and {BenZvi}, S. and {Bianchi}, D. and {Blake}, C. and {Brieden}, S. and {Brodzeller}, A. and {Brooks}, D. and {Buckley-Geer}, E. and {Burtin}, E. and {Calderon}, R. and {Canning}, R. and {Rosell}, A. Carnero and {Carrilho}, P. and {Casas}, L. and {Castander}, F.~J. and {Charles}, M. and {Chaussidon}, E. and {Chaves-Montero}, J. and {Chebat}, D. and {Chen}, X. and {Claybaugh}, T. and {Cole}, S. and {Cooper}, A.~P. and {Cuceu}, A. and {Dawson}, K.~S. and {de la Macorra}, A. and {de Mattia}, A. and {Deiosso}, N. and {Della Costa}, J. and {Demina}, R. and {Dey}, A. and {Dey}, B. and {Ding}, Z. and {Doel}, P. and {Edelstein}, J. and {Eisenstein}, D.~J. and {Elbers}, W. and {Fagrelius}, P. and {Fanning}, K. and {Fern{\'a}ndez-Garc{\'\i}a}, E. and {Ferraro}, S. and {Font-Ribera}, A. and {Forero-Romero}, J.~E. and {Frenk}, C.~S. and {Garcia-Quintero}, C. and {Garrison}, L.~H. and {Gazta{\~n}aga}, E. and {Gil-Mar{\'\i}n}, H. and {Gontcho A Gontcho}, S. and {Gonzalez}, D. and {Gonzalez-Morales}, A.~X. and {Gordon}, C. and {Green}, D. and {Gutierrez}, G. and {Guy}, J. and {Hadzhiyska}, B. and {Hahn}, C. and {He}, S. and {Herbold}, M. and {Herrera-Alcantar}, H.~K. and {Ho}, M.-F. and {Honscheid}, K. and {Howlett}, C. and {Huterer}, D. and {Ishak}, M. and {Juneau}, S. and {Kamble}, N.~V. and {Kara{\c{c}}ayl{\i}}, N.~G. and {Kehoe}, R. and {Kent}, S. and {Kim}, A.~G. and {Kirkby}, D. and {Kisner}, T. and {Koposov}, S.~E. and {Kremin}, A. and {Krolewski}, A. and {Lahav}, O. and {Lamman}, C. and {Landriau}, M. and {Lang}, D. and {Lasker}, J. and {Le Goff}, J.~M. and {Le Guillou}, L. and {Leauthaud}, A. and {Levi}, M.~E. and {Li}, Q. and {Li}, T.~S. and {Lodha}, K. and {Lokken}, M. and {Lozano-Rodr{\'\i}guez}, F. and {Magneville}, C. and {Manera}, M. and {Martini}, P. and {Matthewson}, W.~L. and {Meisner}, A. and {Mena-Fern{\'a}ndez}, J. and {Menegas}, A. and {Mergulh{\~a}o}, T. and {Miquel}, R. and {Moustakas}, J. and {Mu{\~n}oz-Guti{\'e}rrez}, A. and {Mu{\~n}oz-Santos}, D. and {Myers}, A.~D. and {Nadathur}, S. and {Naidoo}, K. and {Napolitano}, L. and {Newman}, J.~A. and {Niz}, G. and {Noriega}, H.~E. and {Paillas}, E. and {Palanque-Delabrouille}, N. and {Pan}, J. and {Peacock}, J.~A. and {Pellejero Ibanez}, M. and {Percival}, W.~J. and {P{\'e}rez-Fern{\'a}ndez}, A. and {P{\'e}rez-R{\`a}fols}, I. and {Pieri}, M.~M. and {Poppett}, C. and {Prada}, F. and {Rabinowitz}, D. and {Raichoor}, A. and {Ram{\'\i}rez-P{\'e}rez}, C. and {Rashkovetskyi}, M. and {Ravoux}, C. and {Rich}, J. and {Rocher}, A. and {Rockosi}, C. and {Rohlf}, J. and {Rom{\'a}n-Herrera}, J.~O. and {Ross}, A.~J. and {Rossi}, G. and {Ruggeri}, R. and {Ruhlmann-Kleider}, V. and {Samushia}, L. and {Sanchez}, E. and {Sanders}, N. and {Schlegel}, D. and {Schubnell}, M. and {Seo}, H. and {Shafieloo}, A. and {Sharples}, R. and {Silber}, J. and {Sinigaglia}, F. and {Sprayberry}, D. and {Tan}, T. and {Tarl{\'e}}, G. and {Taylor}, P. and {Turner}, W. and {Ure{\~n}a-L{\'o}pez}, L.~A. and {Vaisakh}, R. and {Valdes}, F. and {Valogiannis}, G. and {Vargas-Maga{\~n}a}, M. and {Verde}, L. and {Walther}, M. and {Weaver}, B.~A. and {Weinberg}, D.~H. and {White}, M. and {Wolfson}, M. and {Y{\`e}che}, C. and {Yu}, J. and {Zaborowski}, E.~A. and {Zarrouk}, P. and {Zhai}, Z. and {Zhang}, H. and {Zhao}, C. and {Zhao}, G.~B. and {Zhou}, R. and {Zou}, H. and {DESI Collaboration}},
        title = "{DESI DR2 results. II. Measurements of baryon acoustic oscillations and cosmological constraints}",
      journal = {\prd},
         year = 2025,
        month = oct,
       volume = {112},
       number = {8},
          eid = {083515},
        pages = {083515},
          doi = {10.1103/tr6y-kpc6},
archivePrefix = {arXiv},
       eprint = {2503.14738},
 primaryClass = {astro-ph.CO},
       adsurl = {https://ui.adsabs.harvard.edu/abs/2025PhRvD.112h3515A}
}

@ARTICLE{2022JHEAp..34...49A,
       author = {{Abdalla}, Elcio and {Abell{\'a}n}, Guillermo Franco and {Aboubrahim}, Amin and {Agnello}, Adriano and {Akarsu}, {\"O}zg{\"u}r and {Akrami}, Yashar and {Alestas}, George and {Aloni}, Daniel and {Amendola}, Luca and {Anchordoqui}, Luis A. and {Anderson}, Richard I. and {Arendse}, Nikki and {Asgari}, Marika and {Ballardini}, Mario and {Barger}, Vernon and {Basilakos}, Spyros and {Batista}, Ronaldo C. and {Battistelli}, Elia S. and {Battye}, Richard and {Benetti}, Micol and {Benisty}, David and {Berlin}, Asher and {de Bernardis}, Paolo and {Berti}, Emanuele and {Bidenko}, Bohdan and {Birrer}, Simon and {Blakeslee}, John P. and {Boddy}, Kimberly K. and {Bom}, Clecio R. and {Bonilla}, Alexander and {Borghi}, Nicola and {Bouchet}, Fran{\c{c}}ois R. and {Braglia}, Matteo and {Buchert}, Thomas and {Buckley-Geer}, Elizabeth and {Calabrese}, Erminia and {Caldwell}, Robert R. and {Camarena}, David and {Capozziello}, Salvatore and {Casertano}, Stefano and {Chen}, Geoff C.-F. and {Chluba}, Jens and {Chen}, Angela and {Chen}, Hsin-Yu and {Chudaykin}, Anton and {Cicoli}, Michele and {Copi}, Craig J. and {Courbin}, Fred and {Cyr-Racine}, Francis-Yan and {Czerny}, Bo{\.z}ena and {Dainotti}, Maria and {D'Amico}, Guido and {Davis}, Anne-Christine and {de Cruz P{\'e}rez}, Javier and {de Haro}, Jaume and {Delabrouille}, Jacques and {Denton}, Peter B. and {Dhawan}, Suhail and {Dienes}, Keith R. and {Di Valentino}, Eleonora and {Du}, Pu and {Eckert}, Dominique and {Escamilla-Rivera}, Celia and {Fert{\'e}}, Agn{\`e}s and {Finelli}, Fabio and {Fosalba}, Pablo and {Freedman}, Wendy L. and {Frusciante}, Noemi and {Gazta{\~n}aga}, Enrique and {Giar{\`e}}, William and {Giusarma}, Elena and {G{\'o}mez-Valent}, Adri{\`a} and {Handley}, Will and {Harrison}, Ian and {Hart}, Luke and {Hazra}, Dhiraj Kumar and {Heavens}, Alan and {Heinesen}, Asta and {Hildebrandt}, Hendrik and {Hill}, J. Colin and {Hogg}, Natalie B. and {Holz}, Daniel E. and {Hooper}, Deanna C. and {Hosseininejad}, Nikoo and {Huterer}, Dragan and {Ishak}, Mustapha and {Ivanov}, Mikhail M. and {Jaffe}, Andrew H. and {Jang}, In Sung and {Jedamzik}, Karsten and {Jimenez}, Raul and {Joseph}, Melissa and {Joudaki}, Shahab and {Kamionkowski}, Marc and {Karwal}, Tanvi and {Kazantzidis}, Lavrentios and {Keeley}, Ryan E. and {Klasen}, Michael and {Komatsu}, Eiichiro and {Koopmans}, L{\'e}on V.~E. and {Kumar}, Suresh and {Lamagna}, Luca and {Lazkoz}, Ruth and {Lee}, Chung-Chi and {Lesgourgues}, Julien and {Levi Said}, Jackson and {Lewis}, Tiffany R. and {L'Huillier}, Benjamin and {Lucca}, Matteo and {Maartens}, Roy and {Macri}, Lucas M. and {Marfatia}, Danny and {Marra}, Valerio and {Martins}, Carlos J.~A.~P. and {Masi}, Silvia and {Matarrese}, Sabino and {Mazumdar}, Arindam and {Melchiorri}, Alessandro and {Mena}, Olga and {Mersini-Houghton}, Laura and {Mertens}, James and {Milakovi{\'c}}, Dinko and {Minami}, Yuto and {Miranda}, Vivian and {Moreno-Pulido}, Cristian and {Moresco}, Michele and {Mota}, David F. and {Mottola}, Emil and {Mozzon}, Simone and {Muir}, Jessica and {Mukherjee}, Ankan and {Mukherjee}, Suvodip and {Naselsky}, Pavel and {Nath}, Pran and {Nesseris}, Savvas and {Niedermann}, Florian and {Notari}, Alessio and {Nunes}, Rafael C. and {{\'O} Colg{\'a}in}, Eoin and {Owens}, Kayla A. and {{\"O}z{\"u}lker}, Emre and {Pace}, Francesco and {Paliathanasis}, Andronikos and {Palmese}, Antonella and {Pan}, Supriya and {Paoletti}, Daniela and {Perez Bergliaffa}, Santiago E. and {Perivolaropoulos}, Leandros and {Pesce}, Dominic W. and {Pettorino}, Valeria and {Philcox}, Oliver H.~E. and {Pogosian}, Levon and {Poulin}, Vivian and {Poulot}, Gaspard and {Raveri}, Marco and {Reid}, Mark J. and {Renzi}, Fabrizio and {Riess}, Adam G. and {Sabla}, Vivian I. and {Salucci}, Paolo and {Salzano}, Vincenzo and {Saridakis}, Emmanuel N. and {Sathyaprakash}, Bangalore S. and {Schmaltz}, Martin and {Sch{\"o}neberg}, Nils and {Scolnic}, Dan and {Sen}, Anjan A. and {Sehgal}, Neelima and {Shafieloo}, Arman and {Sheikh-Jabbari}, M.~M. and {Silk}, Joseph and {Silvestri}, Alessandra and {Skara}, Foteini and {Sloth}, Martin S. and {Soares-Santos}, Marcelle and {Sol{\`a} Peracaula}, Joan and {Songsheng}, Yu-Yang and {Soriano}, Jorge F. and {Staicova}, Denitsa and {Starkman}, Glenn D. and {Szapudi}, Istv{\'a}n and {Teixeira}, Elsa M. and {Thomas}, Brooks and {Treu}, Tommaso and {Trott}, Emery and {van de Bruck}, Carsten and {Vazquez}, J. Alberto and {Verde}, Licia and {Visinelli}, Luca and {Wang}, Deng and {Wang}, Jian-Min and {Wang}, Shao-Jiang and {Watkins}, Richard and {Watson}, Scott and {Webb}, John K. and {Weiner}, Neal and {Weltman}, Amanda and {Witte}, Samuel J. and {Wojtak}, Rados{\l}aw and {Yadav}, Anil Kumar},
        title = "{Cosmology intertwined: A review of the particle physics, astrophysics, and cosmology associated with the cosmological tensions and anomalies}",
      journal = {Journal of High Energy Astrophysics},
         year = 2022,
        month = jun,
       volume = {34},
        pages = {49-211},
          doi = {10.1016/j.jheap.2022.04.002},
archivePrefix = {arXiv},
       eprint = {2203.06142},
 primaryClass = {astro-ph.CO},
       adsurl = {https://ui.adsabs.harvard.edu/abs/2022JHEAp..34...49A}
}

@ARTICLE{2022LRR....25....6M,
       author = {{Moresco}, Michele and {Amati}, Lorenzo and {Amendola}, Luca and {Birrer}, Simon and {Blakeslee}, John P. and {Cantiello}, Michele and {Cimatti}, Andrea and {Darling}, Jeremy and {Della Valle}, Massimo and {Fishbach}, Maya and {Grillo}, Claudio and {Hamaus}, Nico and {Holz}, Daniel and {Izzo}, Luca and {Jimenez}, Raul and {Lusso}, Elisabeta and {Meneghetti}, Massimo and {Piedipalumbo}, Ester and {Pisani}, Alice and {Pourtsidou}, Alkistis and {Pozzetti}, Lucia and {Quartin}, Miguel and {Risaliti}, Guido and {Rosati}, Piero and {Verde}, Licia},
        title = "{Unveiling the Universe with emerging cosmological probes}",
      journal = {Living Reviews in Relativity},
         year = 2022,
        month = dec,
       volume = {25},
       number = {1},
          eid = {6},
        pages = {6},
          doi = {10.1007/s41114-022-00040-z},
archivePrefix = {arXiv},
       eprint = {2201.07241},
 primaryClass = {astro-ph.CO},
       adsurl = {https://ui.adsabs.harvard.edu/abs/2022LRR....25....6M}
}

@ARTICLE{2023ARNPS..73..153K,
       author = {{Kamionkowski}, Marc and {Riess}, Adam G.},
        title = "{The Hubble Tension and Early Dark Energy}",
      journal = {Annual Review of Nuclear and Particle Science},
         year = 2023,
        month = sep,
       volume = {73},
        pages = {153-180},
          doi = {10.1146/annurev-nucl-111422-024107},
archivePrefix = {arXiv},
       eprint = {2211.04492},
 primaryClass = {astro-ph.CO},
       adsurl = {https://ui.adsabs.harvard.edu/abs/2023ARNPS..73..153K}
}

@ARTICLE{2021CQGra..38o3001D,
       author = {{Di Valentino}, Eleonora and {Mena}, Olga and {Pan}, Supriya and {Visinelli}, Luca and {Yang}, Weiqiang and {Melchiorri}, Alessandro and {Mota}, David F. and {Riess}, Adam G. and {Silk}, Joseph},
        title = "{In the realm of the Hubble tension-a review of solutions}",
      journal = {Classical and Quantum Gravity},
         year = 2021,
        month = jul,
       volume = {38},
       number = {15},
          eid = {153001},
        pages = {153001},
          doi = {10.1088/1361-6382/ac086d},
archivePrefix = {arXiv},
       eprint = {2103.01183},
 primaryClass = {astro-ph.CO},
       adsurl = {https://ui.adsabs.harvard.edu/abs/2021CQGra..38o3001D}
}

@ARTICLE{2002ApJ...573...37J,
       author = {{Jimenez}, Raul and {Loeb}, Abraham},
        title = "{Constraining Cosmological Parameters Based on Relative Galaxy Ages}",
      journal = {\apj},
         year = 2002,
        month = jul,
       volume = {573},
       number = {1},
        pages = {37-42},
          doi = {10.1086/340549},
archivePrefix = {arXiv},
       eprint = {astro-ph/0106145},
 primaryClass = {astro-ph},
       adsurl = {https://ui.adsabs.harvard.edu/abs/2002ApJ...573...37J}
}

@ARTICLE{2011JCAP...03..045M,
       author = {{Moresco}, Michele and {Jimenez}, Raul and {Cimatti}, Andrea and {Pozzetti}, Lucia},
        title = "{Constraining the expansion rate of the Universe using low-redshift ellipticals as cosmic chronometers}",
      journal = {\jcap},
         year = 2011,
        month = mar,
       volume = {2011},
       number = {3},
          eid = {045},
        pages = {045},
          doi = {10.1088/1475-7516/2011/03/045},
archivePrefix = {arXiv},
       eprint = {1010.0831},
 primaryClass = {astro-ph.CO},
       adsurl = {https://ui.adsabs.harvard.edu/abs/2011JCAP...03..045M}
}

@ARTICLE{2022JHEAp..36...27V,
       author = {{Vagnozzi}, Sunny and {Pacucci}, Fabio and {Loeb}, Abraham},
        title = "{Implications for the Hubble tension from the ages of the oldest astrophysical objects}",
      journal = {Journal of High Energy Astrophysics},
         year = 2022,
        month = nov,
       volume = {36},
        pages = {27-35},
          doi = {10.1016/j.jheap.2022.07.004},
archivePrefix = {arXiv},
       eprint = {2105.10421},
 primaryClass = {astro-ph.CO},
       adsurl = {https://ui.adsabs.harvard.edu/abs/2022JHEAp..36...27V}
}

@ARTICLE{2003astro.ph..1623E,
       author = {{Eisenstein}, Daniel},
        title = "{Large-Scale Structure and Future Surveys}",
      journal = {arXiv e-prints},
         year = 2003,
        month = jan,
          eid = {astro-ph/0301623},
        pages = {astro-ph/0301623},
          doi = {10.48550/arXiv.astro-ph/0301623},
archivePrefix = {arXiv},
       eprint = {astro-ph/0301623},
 primaryClass = {astro-ph},
       adsurl = {https://ui.adsabs.harvard.edu/abs/2003astro.ph..1623E}
}

@ARTICLE{2003ApJ...594..665B,
       author = {{Blake}, Chris and {Glazebrook}, Karl},
        title = "{Probing Dark Energy Using Baryonic Oscillations in the Galaxy Power Spectrum as a Cosmological Ruler}",
      journal = {\apj},
         year = 2003,
        month = sep,
       volume = {594},
       number = {2},
        pages = {665-673},
          doi = {10.1086/376983},
archivePrefix = {arXiv},
       eprint = {astro-ph/0301632},
 primaryClass = {astro-ph},
       adsurl = {https://ui.adsabs.harvard.edu/abs/2003ApJ...594..665B}
}

@ARTICLE{2019MNRAS.490..417C,
       author = {{Carnall}, A.~C. and {McLure}, R.~J. and {Dunlop}, J.~S. and {Cullen}, F. and {McLeod}, D.~J. and {Wild}, V. and {Johnson}, B.~D. and {Appleby}, S. and {Dav{\'e}}, R. and {Amorin}, R. and {Bolzonella}, M. and {Castellano}, M. and {Cimatti}, A. and {Cucciati}, O. and {Gargiulo}, A. and {Garilli}, B. and {Marchi}, F. and {Pentericci}, L. and {Pozzetti}, L. and {Schreiber}, C. and {Talia}, M. and {Zamorani}, G.},
        title = "{The VANDELS survey: the star-formation histories of massive quiescent galaxies at 1.0 < z < 1.3}",
      journal = {\mnras},
         year = 2019,
        month = nov,
       volume = {490},
       number = {1},
        pages = {417-439},
          doi = {10.1093/mnras/stz2544},
archivePrefix = {arXiv},
       eprint = {1903.11082},
 primaryClass = {astro-ph.GA},
       adsurl = {https://ui.adsabs.harvard.edu/abs/2019MNRAS.490..417C}
}

@ARTICLE{1996AJ....112..839C,
       author = {{Cowie}, Lennox L. and {Songaila}, Antoinette and {Hu}, Esther M. and {Cohen}, J.~G.},
        title = "{New Insight on Galaxy Formation and Evolution From Keck Spectroscopy of the Hawaii Deep Fields}",
      journal = {\aj},
         year = 1996,
        month = sep,
       volume = {112},
        pages = {839},
          doi = {10.1086/118058},
archivePrefix = {arXiv},
       eprint = {astro-ph/9606079},
 primaryClass = {astro-ph},
       adsurl = {https://ui.adsabs.harvard.edu/abs/1996AJ....112..839C}
}

@ARTICLE{2014ApJ...788...72G,
       author = {{Gallazzi}, Anna and {Bell}, Eric F. and {Zibetti}, Stefano and {Brinchmann}, Jarle and {Kelson}, Daniel D.},
        title = "{Charting the Evolution of the Ages and Metallicities of Massive Galaxies since z = 0.7}",
      journal = {\apj},
         year = 2014,
        month = jun,
       volume = {788},
       number = {1},
          eid = {72},
        pages = {72},
          doi = {10.1088/0004-637X/788/1/72},
archivePrefix = {arXiv},
       eprint = {1404.5624},
 primaryClass = {astro-ph.GA},
       adsurl = {https://ui.adsabs.harvard.edu/abs/2014ApJ...788...72G}
}

@ARTICLE{2022ApJ...929..131C,
       author = {{Carnall}, Adam C. and {McLure}, Ross J. and {Dunlop}, James S. and {Hamadouche}, Massissilia and {Cullen}, Fergus and {McLeod}, Derek J. and {Begley}, Ryan and {Amorin}, Ricardo and {Bolzonella}, Micol and {Castellano}, Marco and {Cimatti}, Andrea and {Fontanot}, Fabio and {Gargiulo}, Adriana and {Garilli}, Bianca and {Mannucci}, Filippo and {Pentericci}, Laura and {Talia}, Margherita and {Zamorani}, Giovani and {Calabro}, Antonello and {Cresci}, Giovanni and {Hathi}, Nimish P.},
        title = "{The Stellar Metallicities of Massive Quiescent Galaxies at 1.0 < z < 1.3 from KMOS + VANDELS}",
      journal = {\apj},
         year = 2022,
        month = apr,
       volume = {929},
       number = {2},
          eid = {131},
        pages = {131},
          doi = {10.3847/1538-4357/ac5b62},
archivePrefix = {arXiv},
       eprint = {2108.13430},
 primaryClass = {astro-ph.GA},
       adsurl = {https://ui.adsabs.harvard.edu/abs/2022ApJ...929..131C}
}

@ARTICLE{2010MNRAS.404.1775T,
       author = {{Thomas}, Daniel and {Maraston}, Claudia and {Schawinski}, Kevin and {Sarzi}, Marc and {Silk}, Joseph},
        title = "{Environment and self-regulation in galaxy formation}",
      journal = {\mnras},
         year = 2010,
        month = jun,
       volume = {404},
       number = {4},
        pages = {1775-1789},
          doi = {10.1111/j.1365-2966.2010.16427.x},
archivePrefix = {arXiv},
       eprint = {0912.0259},
 primaryClass = {astro-ph.CO},
       adsurl = {https://ui.adsabs.harvard.edu/abs/2010MNRAS.404.1775T}
}

@ARTICLE{2018MNRAS.480.4379C,
       author = {{Carnall}, A.~C. and {McLure}, R.~J. and {Dunlop}, J.~S. and {Dav{\'e}}, R.},
        title = "{Inferring the star formation histories of massive quiescent galaxies with BAGPIPES: evidence for multiple quenching mechanisms}",
      journal = {\mnras},
         year = 2018,
        month = nov,
       volume = {480},
       number = {4},
        pages = {4379-4401},
          doi = {10.1093/mnras/sty2169},
archivePrefix = {arXiv},
       eprint = {1712.04452},
 primaryClass = {astro-ph.GA},
       adsurl = {https://ui.adsabs.harvard.edu/abs/2018MNRAS.480.4379C}
}

@book{hubble1982realm,
  title={The realm of the nebulae},
  author={Hubble, Edwin Powell},
  volume={25},
  year={1982},
  publisher={Yale University Press}
}

@ARTICLE{2013A&A...556A..55I,
       author = {{Ilbert}, O. and {McCracken}, H.~J. and {Le F{\`e}vre}, O. and {Capak}, P. and {Dunlop}, J. and {Karim}, A. and {Renzini}, M.~A. and {Caputi}, K. and {Boissier}, S. and {Arnouts}, S. and {Aussel}, H. and {Comparat}, J. and {Guo}, Q. and {Hudelot}, P. and {Kartaltepe}, J. and {Kneib}, J.~P. and {Krogager}, J.~K. and {Le Floc'h}, E. and {Lilly}, S. and {Mellier}, Y. and {Milvang-Jensen}, B. and {Moutard}, T. and {Onodera}, M. and {Richard}, J. and {Salvato}, M. and {Sanders}, D.~B. and {Scoville}, N. and {Silverman}, J.~D. and {Taniguchi}, Y. and {Tasca}, L. and {Thomas}, R. and {Toft}, S. and {Tresse}, L. and {Vergani}, D. and {Wolk}, M. and {Zirm}, A.},
        title = "{Mass assembly in quiescent and star-forming galaxies since z ≃ 4 from UltraVISTA}",
      journal = {\aap},
         year = 2013,
        month = aug,
       volume = {556},
          eid = {A55},
        pages = {A55},
          doi = {10.1051/0004-6361/201321100},
archivePrefix = {arXiv},
       eprint = {1301.3157},
 primaryClass = {astro-ph.CO},
       adsurl = {https://ui.adsabs.harvard.edu/abs/2013A&A...556A..55I}
}

@ARTICLE{2010A&A...523A..13P,
       author = {{Pozzetti}, L. and {Bolzonella}, M. and {Zucca}, E. and {Zamorani}, G. and {Lilly}, S. and {Renzini}, A. and {Moresco}, M. and {Mignoli}, M. and {Cassata}, P. and {Tasca}, L. and {Lamareille}, F. and {Maier}, C. and {Meneux}, B. and {Halliday}, C. and {Oesch}, P. and {Vergani}, D. and {Caputi}, K. and {Kova{\v{c}}}, K. and {Cimatti}, A. and {Cucciati}, O. and {Iovino}, A. and {Peng}, Y. and {Carollo}, M. and {Contini}, T. and {Kneib}, J.-P. and {Le F{\'e}vre}, O. and {Mainieri}, V. and {Scodeggio}, M. and {Bardelli}, S. and {Bongiorno}, A. and {Coppa}, G. and {de la Torre}, S. and {de Ravel}, L. and {Franzetti}, P. and {Garilli}, B. and {Kampczyk}, P. and {Knobel}, C. and {Le Borgne}, J.-F. and {Le Brun}, V. and {Pell{\`o}}, R. and {Perez Montero}, E. and {Ricciardelli}, E. and {Silverman}, J.~D. and {Tanaka}, M. and {Tresse}, L. and {Abbas}, U. and {Bottini}, D. and {Cappi}, A. and {Guzzo}, L. and {Koekemoer}, A.~M. and {Leauthaud}, A. and {Maccagni}, D. and {Marinoni}, C. and {McCracken}, H.~J. and {Memeo}, P. and {Porciani}, C. and {Scaramella}, R. and {Scarlata}, C. and {Scoville}, N.},
        title = "{zCOSMOS - 10k-bright spectroscopic sample. The bimodality in the galaxy stellar mass function: exploring its evolution with redshift}",
      journal = {\aap},
         year = 2010,
        month = nov,
       volume = {523},
          eid = {A13},
        pages = {A13},
          doi = {10.1051/0004-6361/200913020},
archivePrefix = {arXiv},
       eprint = {0907.5416},
 primaryClass = {astro-ph.CO},
       adsurl = {https://ui.adsabs.harvard.edu/abs/2010A&A...523A..13P}
}

@ARTICLE{2022ApJ...927..164B,
       author = {{Borghi}, Nicola and {Moresco}, Michele and {Cimatti}, Andrea and {Huchet}, Alexandre and {Quai}, Salvatore and {Pozzetti}, Lucia},
        title = "{Toward a Better Understanding of Cosmic Chronometers: Stellar Population Properties of Passive Galaxies at Intermediate Redshift}",
      journal = {\apj},
         year = 2022,
        month = mar,
       volume = {927},
       number = {2},
          eid = {164},
        pages = {164},
          doi = {10.3847/1538-4357/ac3240},
archivePrefix = {arXiv},
       eprint = {2106.14894},
 primaryClass = {astro-ph.GA},
       adsurl = {https://ui.adsabs.harvard.edu/abs/2022ApJ...927..164B}
}

@ARTICLE{2022ApJ...928L...4B,
       author = {{Borghi}, Nicola and {Moresco}, Michele and {Cimatti}, Andrea},
        title = "{Toward a Better Understanding of Cosmic Chronometers: A New Measurement of H(z) at z   0.7}",
      journal = {\apjl},
         year = 2022,
        month = mar,
       volume = {928},
       number = {1},
          eid = {L4},
        pages = {L4},
          doi = {10.3847/2041-8213/ac3fb2},
archivePrefix = {arXiv},
       eprint = {2110.04304},
 primaryClass = {astro-ph.CO},
       adsurl = {https://ui.adsabs.harvard.edu/abs/2022ApJ...928L...4B}
}

@ARTICLE{2023A&A...679A..96T,
       author = {{Tomasetti}, E. and {Moresco}, M. and {Borghi}, N. and {Jiao}, K. and {Cimatti}, A. and {Pozzetti}, L. and {Carnall}, A.~C. and {McLure}, R.~J. and {Pentericci}, L.},
        title = "{A new measurement of the expansion history of the Universe at z = 1.26 with cosmic chronometers in VANDELS}",
      journal = {\aap},
         year = 2023,
        month = nov,
       volume = {679},
          eid = {A96},
        pages = {A96},
          doi = {10.1051/0004-6361/202346992},
archivePrefix = {arXiv},
       eprint = {2305.16387},
 primaryClass = {astro-ph.CO},
       adsurl = {https://ui.adsabs.harvard.edu/abs/2023A&A...679A..96T}
}

@ARTICLE{2023ApJS..265...48J,
       author = {{Jiao}, Kang and {Borghi}, Nicola and {Moresco}, Michele and {Zhang}, Tong-Jie},
        title = "{New Observational H(z) Data from Full-spectrum Fitting of Cosmic Chronometers in the LEGA-C Survey}",
      journal = {\apjs},
         year = 2023,
        month = apr,
       volume = {265},
       number = {2},
          eid = {48},
        pages = {48},
          doi = {10.3847/1538-4365/acbc77},
archivePrefix = {arXiv},
       eprint = {2205.05701},
 primaryClass = {astro-ph.CO},
       adsurl = {https://ui.adsabs.harvard.edu/abs/2023ApJS..265...48J}
}

@ARTICLE{2023JCAP...11..051E,
       author = {{Escamilla}, Luis A. and {Akarsu}, {\"O}zg{\"u}r and {Di Valentino}, Eleonora and {Vazquez}, J. Alberto},
        title = "{Model-independent reconstruction of the interacting dark energy kernel: Binned and Gaussian process}",
      journal = {\jcap},
         year = 2023,
        month = nov,
       volume = {2023},
       number = {11},
          eid = {051},
        pages = {051},
          doi = {10.1088/1475-7516/2023/11/051},
archivePrefix = {arXiv},
       eprint = {2305.16290},
 primaryClass = {astro-ph.CO},
       adsurl = {https://ui.adsabs.harvard.edu/abs/2023JCAP...11..051E}
}

@ARTICLE{2026MNRAS.547ag430L,
       author = {{Lei}, Lei and {Wang}, Ze-Fan and {Wang}, Tong-Lin and {Wang}, Yi-Ying and {Yuan}, Guan-Wen and {Lin}, Wei-Long and {Fan}, Yi-Zhong},
        title = "{Stringent constraint on the CCC + TL cosmology with H(z) measurements}",
      journal = {\mnras},
         year = 2026,
        month = apr,
       volume = {547},
       number = {3},
          eid = {stag430},
        pages = {stag430},
          doi = {10.1093/mnras/stag430},
archivePrefix = {arXiv},
       eprint = {2508.04277},
 primaryClass = {astro-ph.CO},
       adsurl = {https://ui.adsabs.harvard.edu/abs/2026MNRAS.547ag430L}
}

@ARTICLE{2018ApJ...868...84M,
       author = {{Moresco}, Michele and {Jimenez}, Raul and {Verde}, Licia and {Pozzetti}, Lucia and {Cimatti}, Andrea and {Citro}, Annalisa},
        title = "{Setting the Stage for Cosmic Chronometers. I. Assessing the Impact of Young Stellar Populations on Hubble Parameter Measurements}",
      journal = {\apj},
         year = 2018,
        month = dec,
       volume = {868},
       number = {2},
          eid = {84},
        pages = {84},
          doi = {10.3847/1538-4357/aae829},
archivePrefix = {arXiv},
       eprint = {1804.05864},
 primaryClass = {astro-ph.CO},
       adsurl = {https://ui.adsabs.harvard.edu/abs/2018ApJ...868...84M}
}

@ARTICLE{2005PhRvD..71l3001S,
       author = {{Simon}, Joan and {Verde}, Licia and {Jimenez}, Raul},
        title = "{Constraints on the redshift dependence of the dark energy potential}",
      journal = {\prd},
         year = 2005,
        month = jun,
       volume = {71},
       number = {12},
          eid = {123001},
        pages = {123001},
          doi = {10.1103/PhysRevD.71.123001},
archivePrefix = {arXiv},
       eprint = {astro-ph/0412269},
 primaryClass = {astro-ph},
       adsurl = {https://ui.adsabs.harvard.edu/abs/2005PhRvD..71l3001S}
}

@article{Zhang_2014,
doi = {10.1088/1674-4527/14/10/002},
url = {https://doi.org/10.1088/1674-4527/14/10/002},
year = {2014},
month = {oct},
publisher = {},
volume = {14},
number = {10},
pages = {1221},
author = {Zhang, Cong and Zhang, Han and Yuan, Shuo and Liu, Siqi and Zhang, Tong-Jie and Sun, Yan-Chun},
title = {Four new observational H(z) data from luminous red galaxies in the Sloan Digital Sky Survey data release seven},
journal = {Research in Astronomy and Astrophysics}
}

@ARTICLE{2013A&A...558A..61M,
       author = {{Moresco}, M. and {Pozzetti}, L. and {Cimatti}, A. and {Zamorani}, G. and {Bolzonella}, M. and {Lamareille}, F. and {Mignoli}, M. and {Zucca}, E. and {Lilly}, S.~J. and {Carollo}, C.~M. and {Contini}, T. and {Kneib}, J.-P. and {Le F{\`e}vre}, O. and {Mainieri}, V. and {Renzini}, A. and {Scodeggio}, M. and {Bardelli}, S. and {Bongiorno}, A. and {Caputi}, K. and {Cucciati}, O. and {de la Torre}, S. and {de Ravel}, L. and {Franzetti}, P. and {Garilli}, B. and {Iovino}, A. and {Kampczyk}, P. and {Knobel}, C. and {Kova{\v{c}}}, K. and {Le Borgne}, J.-F. and {Le Brun}, V. and {Maier}, C. and {Pell{\'o}}, R. and {Peng}, Y. and {Perez-Montero}, E. and {Presotto}, V. and {Silverman}, J.~D. and {Tanaka}, M. and {Tasca}, L. and {Tresse}, L. and {Vergani}, D. and {Barnes}, L. and {Bordoloi}, R. and {Cappi}, A. and {Diener}, C. and {Koekemoer}, A.~M. and {Le Floc'h}, E. and {L{\'o}pez-Sanjuan}, C. and {McCracken}, H.~J. and {Nair}, P. and {Oesch}, P. and {Scarlata}, C. and {Scoville}, N. and {Welikala}, N.},
        title = "{Spot the difference. Impact of different selection criteria on observed properties of passive galaxies in zCOSMOS-20k sample}",
      journal = {\aap},
         year = 2013,
        month = oct,
       volume = {558},
          eid = {A61},
        pages = {A61},
          doi = {10.1051/0004-6361/201321797},
archivePrefix = {arXiv},
       eprint = {1305.1308},
 primaryClass = {astro-ph.CO},
       adsurl = {https://ui.adsabs.harvard.edu/abs/2013A&A...558A..61M}
}

@ARTICLE{2019OJAp....2E..10F,
       author = {{Feroz}, Farhan and {Hobson}, Michael P. and {Cameron}, Ewan and {Pettitt}, Anthony N.},
        title = "{Importance Nested Sampling and the MultiNest Algorithm}",
      journal = {The Open Journal of Astrophysics},
         year = 2019,
        month = nov,
       volume = {2},
       number = {1},
          eid = {10},
        pages = {10},
          doi = {10.21105/astro.1306.2144},
archivePrefix = {arXiv},
       eprint = {1306.2144},
 primaryClass = {astro-ph.IM},
       adsurl = {https://ui.adsabs.harvard.edu/abs/2019OJAp....2E..10F}
}

@ARTICLE{2014A&A...564A.125B,
       author = {{Buchner}, J. and {Georgakakis}, A. and {Nandra}, K. and {Hsu}, L. and {Rangel}, C. and {Brightman}, M. and {Merloni}, A. and {Salvato}, M. and {Donley}, J. and {Kocevski}, D.},
        title = "{X-ray spectral modelling of the AGN obscuring region in the CDFS: Bayesian model selection and catalogue}",
      journal = {\aap},
         year = 2014,
        month = apr,
       volume = {564},
          eid = {A125},
        pages = {A125},
          doi = {10.1051/0004-6361/201322971},
archivePrefix = {arXiv},
       eprint = {1402.0004},
 primaryClass = {astro-ph.HE},
       adsurl = {https://ui.adsabs.harvard.edu/abs/2014A&A...564A.125B}
}

@ARTICLE{2023MNRAS.525.3181L,
       author = {{Lange}, Johannes U.},
        title = "{NAUTILUS: boosting Bayesian importance nested sampling with deep learning}",
      journal = {\mnras},
         year = 2023,
        month = oct,
       volume = {525},
       number = {2},
        pages = {3181-3194},
          doi = {10.1093/mnras/stad2441},
archivePrefix = {arXiv},
       eprint = {2306.16923},
 primaryClass = {astro-ph.IM},
       adsurl = {https://ui.adsabs.harvard.edu/abs/2023MNRAS.525.3181L}
}

@ARTICLE{2003MNRAS.344.1000B,
       author = {{Bruzual}, G. and {Charlot}, S.},
        title = "{Stellar population synthesis at the resolution of 2003}",
      journal = {\mnras},
         year = 2003,
        month = oct,
       volume = {344},
       number = {4},
        pages = {1000-1028},
          doi = {10.1046/j.1365-8711.2003.06897.x},
archivePrefix = {arXiv},
       eprint = {astro-ph/0309134},
 primaryClass = {astro-ph},
       adsurl = {https://ui.adsabs.harvard.edu/abs/2003MNRAS.344.1000B}
}

@ARTICLE{2001MNRAS.322..231K,
       author = {{Kroupa}, Pavel},
        title = "{On the variation of the initial mass function}",
      journal = {\mnras},
         year = 2001,
        month = apr,
       volume = {322},
       number = {2},
        pages = {231-246},
          doi = {10.1046/j.1365-8711.2001.04022.x},
archivePrefix = {arXiv},
       eprint = {astro-ph/0009005},
 primaryClass = {astro-ph},
       adsurl = {https://ui.adsabs.harvard.edu/abs/2001MNRAS.322..231K}
}

@ARTICLE{2001MNRAS.323..887C,
       author = {{Charlot}, St{\'e}phane and {Longhetti}, Marcella},
        title = "{Nebular emission from star-forming galaxies}",
      journal = {\mnras},
         year = 2001,
        month = may,
       volume = {323},
       number = {4},
        pages = {887-903},
          doi = {10.1046/j.1365-8711.2001.04260.x},
archivePrefix = {arXiv},
       eprint = {astro-ph/0101097},
 primaryClass = {astro-ph},
       adsurl = {https://ui.adsabs.harvard.edu/abs/2001MNRAS.323..887C}
}

@article{hartigan1985dip,
  title={The dip test of unimodality},
  author={Hartigan, John A and Hartigan, Pamela M},
  journal={The annals of Statistics},
  pages={70--84},
  year={1985},
  publisher={JSTOR}
}

@article{hartigan1985algorithm,
  title={Algorithm AS 217: Computation of the dip statistic to test for unimodality},
  author={Hartigan, PM},
  journal={Journal of the Royal Statistical Society. Series C (Applied Statistics)},
  volume={34},
  number={3},
  pages={320--325},
  year={1985},
  publisher={JSTOR}
}
\bibliographystyle{aasjournalv7}


\end{CJK*}
\end{document}